\documentclass[a4paper,11pt]{article}
\usepackage{jheppub} 
\usepackage{braket}
\usepackage{comment}
\usepackage{tikz}
\usepackage{xcolor}
\usepackage{hhline}
\usepackage{slashed}
\usepackage{graphics}
\usepackage{longtable}
\usepackage[compat=1.1.0]{tikz-feynman}
\usepackage{pdflscape}
\usepackage{colortbl}
\usepackage{caption}
\usepackage{subcaption}
\usepackage{tensor}
\usepackage{ytableau}
\usepackage{bm}
\usepackage{graphicx}
\usepackage{bm}

\newcommand{\im}{\mathrm{i}}

\renewcommand{\vec}[1]{\mathbf{#1}}

\renewcommand{\H}{\mathcal{H}}
\newcommand{\SU}{\mathrm{SU}}
\newcommand{\Sp}{\mathrm{Sp}}
\newcommand{\su}{\mathfrak{su}}
\renewcommand{\sp}{\mathfrak{sp}}

\toccontinuoustrue

\DeclareMathAlphabet\mathbfcal{OMS}{cmsy}{b}{n}
\def\[#1\]{\begin{equation}\begin{aligned}#1\end{aligned}\end{equation}}

\title{Deconstructed Weak Isospin from a Symplectic Symmetry}

\author[a]{Alastair Gosnay}
\author[a]{and David J Miller}
\affiliation[a]{School of Physics and Astronomy, University of Glasgow, Glasgow G12 8QQ, United Kingdom}

\emailAdd{a.gosnay.1@research.gla.ac.uk,david.j.miller@glasgow.ac.uk}

\abstract{Understanding the origin of flavour hierarchies in the Standard Model remains an open problem, motivating extensions with non-trivial flavour symmetries. We unify deconstructed weak isospin $\SU(2)_\text{L}^3$ into an $\Sp(6)_\text{L}$ symmetry at a high scale $v_S$. The three generations of Standard Model (SM) left-handed doublets are unified into a single fundamental representation of $\Sp(6)_\text{L}$. In addition to two BSM triplets from the breaking of $\SU(2)_\text{L}^3$, the enlarged symmetry predicts six additional gauge bosons below the unification scale: three $\SU(2)_\text{L}$ triplets and three singlets, which induce flavour transitions in both the quark and lepton sectors, even in the presence of mass-gauge alignment. We derive updated bounds on the intermediate breaking scale $v_{12}$ of $\SU(2)_\text{L}^3$ in the presence of the unification scale $v_S$, mapping exclusions in the $(v_{12}, r)$ parameter space with $r = v_S^2 / v_{12}^2$. The most stringent constraints arise from precision flavour observables involving first- and second-generation transitions, including neutral meson mixing ($K^0 - \bar{K}^0$, $D^0 - \bar{D}^0$), $\mu \to 3e$ and $\mu \to e$ conversion in nuclei. These measurements probe scales well beyond direct collider reach and imply $v_{12} \gtrsim 550 ~\text{TeV}$ for small $r \approx 1$, while for $r\gtrsim100$ the bound relaxes to $v_{12} \gtrsim 150~\text{TeV}$, recovering the limits of the $\SU(2)_\text{L}^3$ model. Additionally we include projections from Mu3e and COMET-I and -II experiments which show promising further reach into the parameter space.
}

\begin{document}
\maketitle

\section{Introduction}

In the Standard Model (SM), the Yukawa coupling matrices between the Higgs and different generations of quarks and charged leptons exhibit large hierarchies in their eigenvalues 
\[
y_u \ll y_c \ll y_t, \quad y_d \ll y_s \ll y_b, \quad y_e \ll y_\mu \ll y_\tau,
\]
and further, the mixing angles required to diagonalise the quark matrices are small and hierarchical. Together these hierarchies form the flavour puzzle of the SM \cite{Feruglio:2015jfa}. Many proposed mechanisms for the origins of these hierarchies involve the introduction of Beyond the Standard Model (BSM) interactions, mediated by yet unobserved new particles which couple differently to the different generations of SM fermions. 

The Yukawa matrices $Y_u$, $Y_d$ and $Y_e$ exhibit an approximate $\mathrm{U}(2)_\text{L}\times \mathrm{U}(2)_\text{R}$ global flavour symmetry under the first two generations of fermions \cite{Barbieri:1995uv, Kagan:2009bn}. Consequently, new physics that approximately respects this symmetry can evade the most stringent flavour constraints, which predominantly arise from processes involving first- and second-generation fermions. In particular, if the couplings of new states to the light generations are sufficiently suppressed, sizeable effects in flavour observables can be avoided, allowing such states to lie at scales as low as a few TeV \cite{Allwicher:2023shc}.

One method of introducing new fields respecting $\mathrm{U}(2)$ is to introduce a new family gauge symmetry which enforces the global family symmetry. These new gauge symmetries are compatible with the SM gauge symmetries and are therefore called horizontal symmetries. The first example of such an approach was introduced by Frogatt and Nielsen \cite{Froggatt:1978nt}, where a gauged $\mathrm{U}(1)_F$ horizontal symmetry is introduced along with heavy vector-like fermions and a `flavon' scalar field to generate effective Yukawa couplings. Other approaches involve the use of non-abelian horizontal symmetries such as $\SU(2)_{l+q}$ \cite{Greljo:2023bix}.

Another possibility is to treat $\mathrm{U}(2)$ as an accidental symmetry arising via flavour deconstruction of the SM gauge interactions. In flavour deconstruction generally, we assume that above the scale of the SM that one of the gauge group factors $G$ of the SM gauge group ($G=\SU(3), ~ \SU(2)_\text{L},~ \text{U}(1)_Y$) has multiple copies which act on the different generations of the SM and is broken to a single flavour-universal copy
\[ \label{eq:diagonal_breaking}
 G_1 \times G_2 \times G_3 \longrightarrow G_{1+2+3}. 
\]
To preserve the approximate $\text{U}(2)$ symmetry we can consider the combined gauge symmetry $G_{1+2}$ (which acts flavour universally on the first two generations) as an intermediate gauge symmetry with breaking scale sitting around an order of magnitude above the electroweak scale. Since the aim is ultimately to have a breaking structure of the form (\ref{eq:diagonal_breaking}), we propose a two-stage breaking of the form 
\[
G_1 \times G_2 \times G_3 \xrightarrow[]{\mathcal{O}(\text{100~TeV})} G_{1+2} \times G_{3} \xrightarrow[]{\mathcal{O}(\text{TeV})} G_{1+2+3},
\]
There has been much recent interest in these types of models  \cite{Fuentes-Martin:2024fpx, Capdevila:2024gki, Covone:2024elw, Isidori:2025rci, Greljo:2024ovt, Barbieri:2023qpf, Davighi:2023iks, Davighi:2023evx, Davighi:2023xqn, Davighi:2025cqx, Lizana:2024jby,FernandezNavarro:2025zmb}. Notably, authors have explored the phenomenology of both deconstructed hypercharge \cite{Davighi:2023evx, FernandezNavarro:2023rhv, FernandezNavarro:2024hnv} and deconstructed weak isospin \cite{Davighi:2023xqn, Capdevila:2024gki}, with the lightest BSM gauge bosons having allowable masses as low as $\mathcal{O}(1\text{TeV}-10\text{TeV})$.

While deconstructed gauge models provide an appealing route to realising approximate flavour symmetries, their ultraviolet origin is typically left unspecified. Embedding such constructions into a simple unified gauge structure can therefore provide a more complete and theoretically motivated framework.

The $\SU(2)^3_\text{L}$ model in particular can be extended by assuming the deconstructed group unifies into a single $\Sp(6)_\text{L}$ symmetry deeper into the UV \cite{Davighi:2022fer}. Since $\SU(2) = \Sp(2)$, with $\epsilon = \im \sigma_2$ playing the role of a symplectic form, $\Sp(6)$ is a very natural choice for enlarging the gauge group of the SM to include flavour. Moreover, $\Sp(6)$ is the smallest simple Lie group that contains $\SU(2)_\text{L}^3$ as a subgroup while admitting a fundamental representation in which the three generations of $\SU(2)_\text{L}$ doublets can be embedded. This minimality provides additional motivation for considering $\Sp(6)_\text{L}$ as a unified gauge structure for flavour. 

Models considering an $\Sp(6)$ symmetry first appeared in the 1980s in the context of $\Sp(6)_\text{L} \times \mathrm{U}(1)_{Y}$ \cite{PhysRevD.30.2011} and $\SU(4)\times \Sp(6)_\text{L} \times \Sp(6)_\text{R}$ \cite{KUO1985641} where it was noticed that the three generations of $\SU(2)_\text{L}$ doublets in the SM could be unified in a single fundamental representation of $\Sp(6)$. Utilising $\SU(2)^3_\text{L}$ as an intermediate symmetry was not explored until recently, again in the context of an $\SU(4)\times \Sp(6)_\text{L} \times \Sp(6)_\text{R}$ model \cite{Davighi:2022fer} where it was shown this intermediate symmetry along with other intermediate stages could generate a CKM matrix compatible with measurements. 

In this work, we extend the findings of \cite{Davighi:2023xqn} by assuming that indeed $\SU(2)^3_\text{L}$ originates from $\Sp(6)_\text{L}$, at scale $v_S$. We therefore consider a model with the breaking pattern 
\[
\Sp(6)_\text{L} &  \xrightarrow[]{v_S} & \SU(2)_{\text{L},1} \times \SU(2)_{\text{L},2} \times \SU(2)_{\text{L},3} \\
 & \xrightarrow[]{v_{12} \sim \mathcal{O}(100 \text{TeV})} & \SU(2)_{\text{L},1+2} \times \SU(2)_{\text{L},3} \\
 & \xrightarrow[]{v_{23} \sim \mathcal{O}(1-10 \text{TeV})} & \SU(2)_{\text{L},1+2+3} \quad (\text{the SM}).
\]
The intermediate breaking of $\SU(2)^3_\text{L}$ to $\SU(2)_{\text{L},1+2+3}$ yields two massive triplets $W_{23}$ and $W_{12}$ with masses constrained to $\gtrsim \mathcal{O}(1 \text{TeV})$ and $\gtrsim \mathcal{O}(100 \text{TeV})$, respectively \cite{Davighi:2023xqn}. The new high scale
symmetry introduces a further three triplets ($\widetilde{W}_{12}$, $\widetilde{W}_{13}$, $\widetilde{W}_{23}$) and three $\SU(2)_\text{L,SM}$ singlets ($Z_{12}$, $Z_{13}$, $Z_{23}$) coming from $\Sp(6)_\text{L}/\SU(2)^3_\text{L}$, with masses of order the high scale $v_S$. Since flavour observables such as meson mixing and lepton flavour violating processes (LFV) already have the potential to probe BSM effects to the scales of hundreds of TeV, we utilise such
observables to constrain the high scale $v_S$.

Since there are a large number of BSM fields in this model, we capture the totality of their effects by first mapping the model onto the SMEFT, providing effective operators with contributions from the three scales $v_{23}$, $v_{12}$ and $v_S$. We then consider current measurements and future planned sensitivities of flavour observables to provide current constraints and future probing potential of the $\Sp(6)_\text{L}$ breaking scale, $v_S$.

\section{The Model}

\subsection{Gauge Group and Field Content} \label{sec:Group&Fields}

We study a model of flavour based on the gauge group
\[ \label{eq:G_uv}
\SU(3)_C \times \Sp(6)_\text{L} \times \text{U}(1)_Y,
\]
where $\SU(3)_C$ and $\text{U}(1)_Y$ are the colour and hypercharge groups of the SM. The weak isospin SU(2)$_\text{L}$ is embedded in a \textit{flavoured} $\Sp(6)_\text{L}$ symmetry, the subgroup of SU(6) satisfying
\[ \label{eq:Sp6}
\Sp(6) & \simeq \{ U \in \SU(6) ~|~ U^\text{T} \Omega U = \Omega \} \\ 
& \simeq \{ U \in \mathcal{M}_{6\times 6}(\mathbb{C}) ~ | ~ U^\dagger U = \mathbb{I}, ~ \text{det} U = 1, ~ U^\text{T} \Omega U = \Omega \},
\]
where $\Omega$ is the symplectic form. The choice of $\Omega$ reflects the choice of basis of the generators and group elements. Since the aim of this work heavily involves the subgroup $\SU(2)^3 \subset \Sp(6)$ it will be convenient to work in the basis where this subgroup is the diagonal subgroup. In this basis we have 
\[ \label{eq:O ega}
\Omega = \begin{pmatrix}
    0 & 1 & 0 & 0 & 0 & 0 \\
    -1 & 0 & 0 & 0 & 0 & 0 \\
    0 & 0 & 0 & 1 & 0 & 0 \\
    0 & 0 & -1 & 0 & 0 & 0 \\
    0 & 0 & 0 & 0 & 0 & 1 \\ 
    0 & 0 & 0 & 0 & -1 & 0 
\end{pmatrix} = \begin{pmatrix}
    \epsilon & 0 & 0 \\
    0 & \epsilon & 0 \\
    0 & 0 & \epsilon
\end{pmatrix},
\]
and for completeness the generators in this basis are given in Appendix \ref{appendix:sp6}. Embedding $\SU(2)_\text{L}$ in $\Sp(6)_{\text{L}}$ is a natural candidate for flavoured gauge symmetry since there is a subgroup $\SU(2) \times \SU(2) \times \SU(2)$ which allows the flavour deconstruction of weak isospin, where we have a separate gauge symmetry for each generation at some high energy scale which can realise the flavour universal $\SU(2)_\text{L}$ of the SM at low energy scales.

The three generations of quark and lepton left-handed $\SU(2_\text{L})$ doublets of the SM ($q_i, l_i$) are combined and lifted to fundamental representations of $\Sp(6)_\text{L}$, the $\vec{6}$ which acts on the vector space $\mathbb{C}^6$:
\[
L = \begin{pmatrix} l_1 \\ l_2 \\l_3 \end{pmatrix} \sim (\vec{1}, \vec{6})_{-1},
\qquad
Q = \begin{pmatrix} q_1 \\ q_2 \\ q_3 \end{pmatrix} \sim (\vec{3}, \vec{6})_{1/3}.
\]
Since we are only altering the structure of gauge symmetries which affects left-chiral fermions, the singlets of $\SU(2)_\text{L}$ in the SM are similarly $\Sp(6)_\text{L}$ singlets. 

The SM Higgs doublet similarly gets embedded in a $\vec{6}$ but requires the introduction of two additional scalar doublets to fillout the representation resulting in a three Higgs doublet model
\[
\mathcal{H} = \begin{pmatrix} H_1 \\ H_2 \\ H_3 \end{pmatrix}  \sim (\vec{1},\vec{6})_{+1}.
\]
Naturally, since we are embedding the Higgs in a flavour enriched multiplet the SM Higgs is aligned with one of the three generations of fermions, and we identify one of the $H_i$ with the SM Higgs doublet. Three generation Higgs models have been explored in detail in \cite{Hartmann:2014ppa, Keus:2013hya, Altmannshofer:2025pjj, Darvishi_2021, Das:2021oik}.

The most suitable alignment for this model is with the 3rd generation \cite{Davighi:2023iks} since the 3rd generation Yukawas are $\mathcal{O}(1)$ and so we denote the SM Higgs as $H_3$. The two other Higgs doublets, aligned with the first and second generation denoted $H_1$ and $H_2$, are assumed to be heavy and do not acquire vevs, and can therefore not contribute to the generation of the masses of the SM fields.  

Fermions couple to the Higgs via terms analogous to those in the SM
\[ \label{eq:UV_yukawas}
\mathcal{L}_{\mathcal{H}\bar{\Psi}\Psi} = \sum_{j=1}^3 y_{e,j} \bar{L} \mathcal{H} e_j + y_{d,j} \bar{Q} \mathcal{H} d_j + y_{u,j} \bar{Q} \tilde{\mathcal{H}} u_j + \text{h.c.}
\]
Here $\tilde{\mathcal{H}} = \Omega \mathcal{H}^*$ which is analogous to the conjugate Higgs doublet in the SM constructed with $\epsilon$. Expanding in terms of the doublets $H_i$ due to the block-nature of the symmetry group
\[ \label{eq:intermediate_yukawas}
\mathcal{L}_{\mathcal{H}\bar{\Psi}\Psi} = \sum_{i,j=1}^3 y_{e,j} \bar{l}_i H_i e_j + y_{d,j} \bar{q}_i H_i d_j + y_{u,j} \bar{q}_i \tilde{H}_i u_j + \text{h.c.}
\]
where as usual $\tilde{H}_i = \epsilon H_i^*$. Since only $\braket{H_3}\neq 0$, the fermion mass matrices only have non-zero entries in the 3rd row and are of the form
\[ \label{eq:yukawa_matrices}
M_{\psi} = v_H \begin{pmatrix}
    0 & 0 & 0 \\ 0 & 0 & 0 \\ y_{\psi,1}  & y_{\psi,2}  & y_{\psi,3} 
\end{pmatrix} \qquad \text{for} \qquad \psi \in \{e, ~d,~ u\},
\]
where $v_H$ is the SM Higgs vev. 

The SM Higgs sector therefore introduces two related issues. First, if only $H_3$ acquires a vev, the renormalisable Yukawa matrices are rank one, so additional structure is required to generate masses for the first- and second-generation fermions. As discussed in Ref.~\cite{Davighi:2023xqn}, this can be described effectively through higher-dimensional Yukawa operators involving the link-field spurions. Second, the vacuum alignment $\langle H_1\rangle=\langle H_2\rangle=0$ is not automatic. For example, if the UV Higgs multiplet has a renormalisable potential of the standard form
\[ \label{eq:higgs_potential}
V = -\mu_\mathcal{H}^2 \mathcal{H}^\dagger \mathcal{H}
+\lambda_\mathcal{H}(\mathcal{H}^\dagger \mathcal{H})^2,
\]
with \(\mu_\mathcal{H}^2>0\), then expanding \(\mathcal{H}\) in terms of the doublets \(H_i\) gives tachyonic mass terms for all three doublets. One possible way to obtain an inert $H_1,H_2$ sector is instead to take $\mu_\mathcal{H}^2<0$ and generate a tachyonic mass term dynamically only for the SM-like Higgs direction through couplings to additional scalars. We introduce two scalar candidates, $S$ and $\Phi$, in the following subsections. However, since this work focuses on the heavy-gauge-boson phenomenology of the $\Sp(6)_\text{L}$ sector, we do not attempt a complete scalar/Yukawa
completion here. We return to these issues in
Sec.~\ref{sec:scalar_yukawa_completion}.

\subsection{Antisymmetric Scalars}

In order to break (\ref{eq:G_uv}) down to the SM, we include two additional scalar fields in the UV, which we denote $S$ and $\Phi$. Both of these fields transform in the $\vec{14}$ representation of $\Sp(6)_\text{L}$, an antisymmetric rank 2 representation which we take to be self-conjugate. Denoting $A \sim \vec{14}$, we can write $A$ as a $6\times6$ matrix which transforms under the gauge group as 
\[
A \longrightarrow U A U^\text{T} \qquad U \in \Sp(6)_\text{L},
\]
where the components of $A$ are subject to the constraint $\text{tr}(\Omega A) = 0$, since the total contraction $A^{ij}\Omega_{ij}$ is a singlet of $\Sp(6)_\text{L}$ and must be subtracted\footnote{This is analogous to the trace being subtracted from symmetric tensors of $\mathrm{SO}(n)$.}. With  $\Omega$ as in (\ref{eq:O ega}), $A$ can conveniently be written in terms of $2 \times 2$ blocks ($\mathcal{A}_{ij}$) of the form
\[
A = \begin{pmatrix}
    \mathcal{A}_{11} & \mathcal{A}_{12} & \mathcal{A}_{13} \\
    -\mathcal{A}_{12}^\text{T} & \mathcal{A}_{22} & \mathcal{A}_{23} \\
    -\mathcal{A}_{13}^\text{T} & -\mathcal{A}_{23}^\text{T} & -(\mathcal{A}_{11} + \mathcal{A}_{22}) \\
\end{pmatrix},
\]
The standard inner product for this representation is the Frobenius inner product, which can be expressed in terms of the block elements by 
\[
\frac{1}{2}\text{tr}(A^\dagger A) = \sum_{i \leq j} \text{tr}(\mathcal{A}^\dagger_{ij} \mathcal{A}_{ij} ) = \sum_{i \leq j} \text{tr} \left|\mathcal{A}_{ij}\right|^2,
\]
with the factor of 1/2 included to account for the antisymmetry of the representation and avoid overcounting. We will often write the second version of this expression for notational convenience, which is the trace of the modulus.

Now we consider embedding scalars within this representation so we can define the necessary components of the fields $S$ and $\Phi$. Since the UV field $S$ obtains a vev, we do not define all the components of the representation since most of its scalar degrees of freedom will give masses to the gauge fields. However, we specify all the components for $\Phi$. 

If we consider the diagonal subgroup $\SU(2)_\text{L}^3 \subset \Sp(6)_\text{L}$, the $\vec{14}$ breaks as 
\[
\vec{14} \rightarrow (\vec{2},\vec{2},\vec{1})\oplus (\vec{2},\vec{1},\vec{2})\oplus (\vec{1},\vec{2},\vec{2})\oplus (\vec{1},\vec{1},\vec{1})\oplus(\vec{1},\vec{1},\vec{1}),
\]
where each element of the tuple is a representation of $\SU(2)_{\text{L},i}$. Due to our choice of basis through (\ref{eq:O ega}), when writing the $\vec{14}$ as a block matrix of $2\times 2$ components, the components correspond to the same representations of the diagonal subgroup as above. Therefore, the UV field $\Phi$ can be expressed as 
\[ \label{eq:phi_components}
\Phi = \begin{pmatrix}
    \frac{1}{\sqrt{6}}\rho_1 - \frac{1}{\sqrt{2}}\rho_2 & \phi_{12} & \phi_{13} \\
    -\phi_{12}^\text{T} & \frac{1}{\sqrt{6}}\rho_1 + \frac{1}{\sqrt{2}}\rho_2 & \phi_{23} \\
    -\phi_{13}^\text{T} & -\phi_{23}^\text{T} & - \sqrt{\frac{2}{3}}\rho_{1} 
\end{pmatrix},
\]
where 
\[
\phi_{12} \sim (\vec{1},[\vec{2},\vec{2}, & \vec{1}])_0, \quad \phi_{13} \sim (\vec{1},[\vec{2},\vec{1},\vec{2}])_0, \quad \phi_{23} \sim (\vec{1},[\vec{1},\vec{2},\vec{2}])_0, \\
& \rho_1 \sim (\vec{1},[\vec{1},\vec{1},\vec{1}])_0, \quad \rho_2 \sim (\vec{1},[\vec{1},\vec{1},\vec{1}])_0.
\]
Here the normalisation is chosen such that 
\[
\frac{1}{2} \text{tr} \left( |\Phi |^2 \right) = \text{tr}\left( |\rho_1 |^2 \right) + \text{tr}\left( |\rho_2 |^2 \right) + \text{tr}\left( |\phi_{12} |^2 \right) + \text{tr}\left( |\phi_{13} |^2 \right) + \text{tr}\left( |\phi_{23} |^2 \right).
\]
We stress that even the singlets here still have a matrix structure, though are proportional to $\epsilon$: $\rho_{i} = \varrho_i \epsilon$ with $\varrho_i$ scalars.

As already mentioned, it is not necessary to specify the components of the UV field $S$, though it would generally take the same form as $\Phi$.

\subsection{Symmetry Breaking Pattern} \label{sec:breaking_pattern}

\begin{figure}[t]
    \centering
    \begin{tikzpicture}
         \node (sp6) at (0,9) {$\Sp(6)_\text{L}$};
         \node (sm) at (0,0) {$\SU(2)_\text{L,SM}$};
         \node (int2) at (0,3) {$\SU(2)_{\text{L},1+2} \times \SU(2)_{\text{L},3}$};
         \node (int1) at (0,6) {$\SU(2)_{\text{L},1} \times \SU(2)_{\text{L},2} \times \SU(2)_{\text{L},3}$};
         \draw[->] (sp6) -- (int1) node[midway, right] {$\braket{S}$} node[midway, left] {SSB (i)} ;
         \draw[->] (int1) -- (int2) node[midway, right] {$\braket{\phi_{12}}$} node[midway, left] {SSB (ii)};
         \draw[->] (int2) -- (sm) node[midway, right] {$\braket{\phi_{23}}$} node[midway, left] {SSB (iii)};
         \draw[line width=0.5mm, -{Latex[length=5mm]}] (6,0) -- (6,10) node[midway, right] {Scale};
         \draw (5.8,1.5) -- (6.2,1.5);
         \node at (5,1.5) {$\mathcal{O}(10 \text{TeV})$};
         \draw (5.8,4.5) -- (6.2,4.5);
         \node at (4.75,4.5) {$\mathcal{O}(100 \text{TeV})$};
    \end{tikzpicture}
    \caption{Breaking pattern from $\Sp(6)_\text{L}$ down to a flavour universal $\SU(2)_\text{L}$. Marked on the right are the lowest orders of magnitude of scales allowed by observation as determined in \cite{Davighi:2023xqn}. }
    \label{fig:breaking_pattern}
\end{figure}
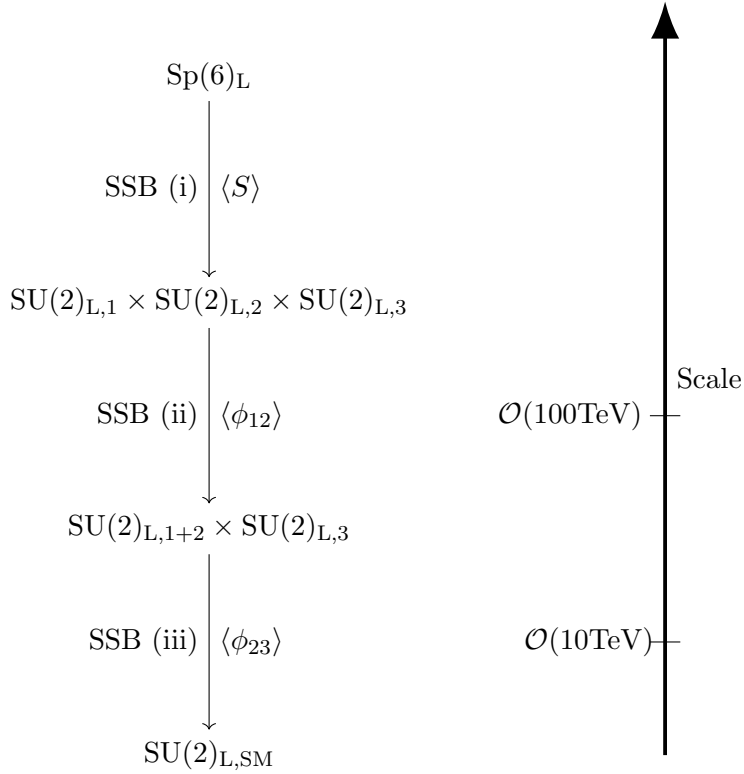

To get from $\Sp(6)_\text{L} \rightarrow \SU(2)_\text{L}$ we go through three phases which are sketched in Figure \ref{fig:breaking_pattern} and described below. 

First, the scalar $S$ acquires a vev at the scale $v_S$, which breaks 
\[
\Sp(6)_\text{L} \longrightarrow \SU(2)_\text{L,1} \times \SU(2)_\text{L,2} \times \SU(2)_\text{L,3}.
\]

Next the bi-doublet $\phi_{12} \subset \Phi$ acquires a vev at scale $v_{12}$ and acts as a link field leaving the diagonal subgroup $\SU(2)_{\text{L,1+2}}$ unbroken:
\[
\SU(2)_\text{L,1} \times \SU(2)_\text{L,2} \longrightarrow \SU(2)_\text{L,1+2}.
\]
This breaking structure results in an accidental (approximate) global $\mathrm{U}(2)$ symmetry\footnote{This symmetry becomes exact as $v_{12} \rightarrow \infty$.} in the first and second generations of fermions below the scale $v_{12}$.

Finally we break to the SM, by breaking $\SU(2)_{\text{L,1+2}}$ and $\SU(2)_{\text{L},3}$ simultaneously through the remaining $\phi_{23} \subset \Phi$ acquiring a vev at the scale $v_{23}$. This leaves the flavour universal $\SU(2)_\text{L}$ of the SM unbroken 
\[
\SU(2)_\text{L,1+2} \times \SU(2)_{\text{L},3} \longrightarrow \SU(2)_\text{L,SM}.
\]

Below the scale $v_S$ the model approaches the phenomenology of \cite{Davighi:2023xqn}, which found lower bounds on the masses of the heavy gauge fields associated with the vevs $v_{12}$ and $v_{23}$. In Figure \ref{fig:breaking_pattern} we mark the orders of magnitude of the vevs with the same bounds found in \cite{Davighi:2023xqn}.

We assume the hierarchy 
\[
v_{23} \ll v_{12} <v_S,
\]
but make no assumption about the scale of $v_S$ relative to $v_{12}$ and assume they could be arbitrarily close if phenomenological considerations allow it. Below we use this hierarchy to compute the masses of the gauge fields associated with the broken symmetries. We refer the reader to Appendix \ref{appendix:sp6} which outlines the structure of the Lie algebra of $\Sp(6)_\text{L}$ for the derivation of any properties stated in the following subsection.

\paragraph{SSB (i)} The first SSB is triggered by the field $S$ getting a vev. Writing the $\Sp(6)_\text{L}$ gauge fields as $\mathcal{W}^\mu_A T_A$, the gauge covariant derivative for $S$ is given by
\[ \label{eq:CovDerS}
D^\mu S = \partial^\mu S - \im g \mathcal{W}^\mu_A \left( T_A S + S T_A^\text{T}  \right),
\]
where $g$ is the Sp(6) gauge coupling. Since $\mathcal{W}^\mu_A T_A \in \mathfrak{sp}(6)$ we can write 
\[ \label{eq:sp(6)elem}
\mathcal{W}^\mu_A T_A =  \begin{pmatrix} x_1^\mu & y_{12}^\mu & y_{13}^\mu \\ y_{12}^{\mu\dagger} &  x_2^\mu & y_{23}^\mu \\ y_{13}^{\mu \dagger} & y_{23}^{\mu \dagger} &  x_3^\mu \end{pmatrix}
\]
where these elements are $2\times 2$ matrices and the indices label the block component (the component of a $3\times 3$ matrix of $2\times 2$ matrices). The indices here also label which $\SU(2)_{\text{L},i}$ factors the field transforms under. That is 
\[
x^\mu_i \rightarrow U_i x^\mu_i U^\dagger_{i}, \quad y^\mu_{ij} \rightarrow U_i y^\mu_{ij} U^\dagger_{j},
\]
with $U_i \in \SU(2)_{\text{L},i}$.

To mediate the breaking, the vev of $S$ is aligned with the components
\[
\braket{S} = \begin{pmatrix}
    \braket{\mathcal{S}} & 0 & 0 \\
    0 & 0 & 0 \\
    0 & 0 & - \braket{\mathcal{S}}
\end{pmatrix} \qquad \text{with} \qquad \braket{\mathcal{S}} =  \begin{pmatrix} 0 & v_S \\ -v_S & 0  \end{pmatrix},
\]
which leave the $\SU(2)^3$ subgroup unbroken, which is associated with the fields $x_i$ in (\ref{eq:sp(6)elem}). We note that this is the same vev used to break $\Sp(6)_\text{L}$ as in \cite{Davighi:2022fer}, though we are working in a different basis. Computing the covariant derivative we find 
\[
\frac{-1}{\im g}D^\mu \braket{S} &= \begin{pmatrix} x^\mu_1 & y^\mu_{12} & y^\mu_{13} \\ y^{\mu \dagger}_{12} &  x^\mu_2 & y^\mu_{23} \\ y_{13}^{\mu \dagger} & y_{23}^{\mu \dagger} &  x^\mu_3 \end{pmatrix} \begin{pmatrix}
    \braket{\mathcal{S}} & 0 & 0 \\
    0 & 0 & 0 \\
    0 & 0 & - \braket{\mathcal{S}}
\end{pmatrix} +
\begin{pmatrix}
    \braket{\mathcal{S}} & 0 & 0 \\
    0 & 0 & 0 \\
    0 & 0 & - \braket{\mathcal{S}}
\end{pmatrix}
 \begin{pmatrix} 
 x_1^{\mu \text{T}} & y_{12}^{\mu *} & y_{13}^{\mu *} \\ 
 y_{12}^{\mu \text{T}} &  x_2^{\mu \text{T}} & y_{23}^{\mu *} \\
 y_{13}^{\mu \text{T}} & y_{23}^{\mu \text{T}} &  x_3^{\mu \text{T}} \end{pmatrix} \\
 &= \begin{pmatrix}
    x^\mu_1 \braket{\mathcal{S}} + \braket{\mathcal{S}} x_1^{\mu \text{T}} && \braket{\mathcal{S}} y_{12}^{\mu *} && - y^\mu _{13} \braket{\mathcal{S}}+\braket{\mathcal{S}} y_{13}^{\mu *} \\
    y_{12}^{\mu \dagger} \braket{\mathcal{S}} && 0 && - y^\mu_{23} \braket{\mathcal{S}} \\
    y_{13}^{\mu \dagger} \braket{\mathcal{S}} - \braket{\mathcal{S}} y_{13}^{\mu \text{T}} && -\braket{\mathcal{S}} y_{23}^{\mu \text{T}} && - x_3^\mu \braket{\mathcal{S}} - \braket{\mathcal{S}} x_3^{\mu \text{T}}
 \end{pmatrix}.
\]
Notice that we can write $\braket{\mathcal{S}} = v_S \epsilon$ and, as derived in Appendix \ref{appendix:sp6}, the elements $x_i$ and $y_{ij}$ obey the following relations involving $\epsilon$
\[
\epsilon x^\mu_i  = -x^{\mu \text{T}}_i \epsilon, \qquad \epsilon y^\mu_{ij} = -y_{ij}^{\mu *} \epsilon.
\]
Therefore, the covariant derivative acting on the vev becomes
\[
D^\mu \braket{S} &= - \im g v_S \begin{pmatrix}
    x^\mu_1 \epsilon + \epsilon x_1^{\mu\text{T}} && \epsilon y_{12}^{\mu *} && - y^\mu_{13} \epsilon+\epsilon y_{13}^{\mu *} \\
    y_{12}^{\mu \dagger} \epsilon && 0 && - y^\mu_{23} \epsilon \\
    y_{13}^{\mu \dagger} \epsilon - \epsilon y_{13}^{\mu \text{T}} && -\epsilon y_{23}^{\mu \text{T}} && - x^\mu_3 \epsilon - \epsilon x_3^{\mu \text{T}}
 \end{pmatrix} \\
&=- \im g v_S 
 \begin{pmatrix}
     0 && - y^\mu_{12} \epsilon && -2 y^\mu_{13} \epsilon \\
     (y^\mu_{12} \epsilon)^\text{T} && 0 && -y^\mu_{23} \epsilon \\
     (2 y^\mu_{13} \epsilon)^\text{T} && (y_{23}^\mu \epsilon)^\text{T} && 0 
 \end{pmatrix}, \\
\]
which preserves the antisymmetric structure of $S$ as expected. Now we can compute the kinetic term for $S$ which yields the masses of the heavy states, giving
\[ \label{eq:S-kinetic-term}
\frac{1}{2} \text{tr}  ( D_\mu \braket{S})^\dagger D^\mu \braket{S} ) =  g^2 v^2_S \left( \text{tr} |y_{12}|^2  + 4 \text{tr}|y_{13}|^2 + \text{tr} |y_{23}|^2 \right).
\]
Under the unbroken subgroup $\SU(2)_\text{L}^3$ these fields are bi-doublets 
$y_{12} \sim (\vec{2},\vec{2},\vec{1})$, $y_{13} \sim (\vec{2},\vec{1},\vec{2})$ and $y_{23} \sim (\vec{1},\vec{2},\vec{2})$.
We note that $\epsilon$ has dropped out of these expressions due to its orthogonality, and we have suppressed Lorentz indices where convenient.

\paragraph{SSB (ii)-(iii)} The subsequent two SSB stages are mediated by the vevs $\phi_{12}$ and $\phi_{23}$ and since both are embedded in the UV representation $\Phi$, we can compute the kinetic term for $\Phi$ which will contain the mass terms that would arise from considering the kinetic terms of the fields separately. Another reason for considering the kinetic term of both fields through $\Phi$ is that we are making no assumptions about the difference between the two high scales $v_{12}$ and $v_S$. As we will see, the scalar $\phi_{12}$ will have non-negligible contributions to the masses generated in (\ref{eq:S-kinetic-term}). If we only considered $\phi_{12}$ (and $\phi_{23}$) in the context of the $\SU(2)^3_\text{L}$ intermediate symmetry, we would also have to account for the gauge fields which are already massive. Therefore, using $\Phi$ can account for contributions to the masses of all the BSM gauge fields of the model.

Since $\Phi \sim (\vec{1},\vec{14})_0$ in the UV, the covariant derivative is of the same form as (\ref{eq:CovDerS})
\[
D_\mu \Phi = \partial_\mu \Phi - \im g \mathcal{W}^A (T_A \Phi + \Phi T_A).
\]
The vevs of the intermediate fields that mediate the two symmetry breaking steps are 
\[
\braket{\phi_{12}} = \begin{pmatrix}
    0 & v_{12} \\
    -v_{12} & 0 
\end{pmatrix} = v_{12} \epsilon, \quad  \braket{\phi_{23}} = \begin{pmatrix}
    0 & v_{23} \\
    -v_{23} & 0 
\end{pmatrix} = v_{23} \epsilon,
\]
so the vev of the UV field $\Phi$ is given by
\[
\braket{\Phi} = \begin{pmatrix}
    0 & \braket{\phi_{12}} & 0 \\
    -\braket{\phi_{12}}^\text{T} & 0 & \braket{\phi_{23}} \\
    0 & -\braket{\phi_{23}}^\text{T} & 0 
\end{pmatrix}.
\]
Just as for $S$, we compute the kinetic term of the vev which yields the mass terms 
\[ \label{eq:Phi_kinetic_term}
\frac{1}{2} \text{tr}\left( (D_\mu \braket{\Phi})^\dagger D^\mu \braket{\Phi} \right) =& 
g^2 \text{tr} | v_{12} (y_{12} - y_{12}^\dagger)|^2 
+ g^2 \text{tr} \left| v_{12}(x_1 - x_2) + v_{23} y_{13}  \right|^2 \\
+& g^2 \text{tr} \left| v_{23} y_{12} - v_{12} y_{23} \right|^2 
+ g^2 \text{tr} \left| v_{12} (y_{12}^\dagger - y_{12}) + v_{23}(y_{23} - y_{23}^\dagger) \right|^2 \\
+& g^2 \text{tr} \left| v_{23} (x_2 - x_3) - v_{12} y_{13} \right|^2 
+ g^2 \text{tr} \left| v_{23} (y_{23}^\dagger - y_{23}) \right|^2. \\
\]
It still remains to rewrite (\ref{eq:Phi_kinetic_term}) and (\ref{eq:S-kinetic-term}) in terms of states which are representations of the unbroken symmetry (which is now $\SU(2)_\text{L,SM}$) whereas the current expressions are in terms of representations of $\SU(2)^3_\text{L}$. 

The resulting states we expect as representations of the SM should only be either triplets or singlets of $\SU(2)_\text{L,SM}$, since the deconstructed factors break down to the diagonal subgroup. Recalling the states of $\SU(2)^3_\text{L}$ are either triplets or bi-doublets, triplets will remain triplets under the diagonal subgroup whereas bi-doublets become \textit{tensor products of doublets} under the diagonal subgroup. For example, in the case of a bi-doublet of $\SU(2)_{\text{L,1}} \times \SU(2)_{\text{L,2}}$, this becomes
\[
(\vec{2},\vec{2},\vec{1}) \longrightarrow \vec{2} \otimes \vec{2} \otimes \vec{1} = \vec{3} \oplus \vec{1}.
\]
In Appendix \ref{appendix:sp6}, we show that we can write
\[
x^\mu_i = w^\mu_{i,I} \tau_I, \quad y^\mu_{ij} = w^\mu_{ij,I} \tau_I + \frac{\im}{2} z^\mu_{ij} \mathbb{I},
\]
where we can explicitly see the splitting of $y^\mu_{ij}$ (the bi doublets) into a triplet and singlet, with $w_I \tau_I \sim \vec{3}$ and $z \sim \vec{1}$. Here we are utilising lower case letters for the fields to identify the fields are not currently in the mass basis. For $y_{ij}$, we have the following expressions for the Hermitian and anti-Hermitian parts 
\[ \label{eq:y-splitting}
y^\mu_{ij} - y_{ij}^{\mu \dagger} = \im z^\mu_{ij} \mathbb{I}, \quad  y^\mu_{ij} + y_{ij}^{\mu \dagger} = 2 w^\mu_{ij,I} \tau_I,
\]
and further have the following inner product relations
\[
\text{tr}|x_i^2| = \frac{1}{2}w_i^2, \quad \text{tr}|y_{ij}|^2 = \frac{1}{2}w_{ij}^2 + \frac{1}{2}z_{ij}^2,
\]
where we have suppressed the $\SU(2)$ vector index $I$ on $w$: $w^2 = \sum_{I} w_I^2$.

In terms of the fields $w^\mu_{i}, ~w^\mu_{ij}$ and $z^\mu_{ij}$, along with the relations (\ref{eq:y-splitting}), the sum of the kinetic terms for $\braket{S}$ (\ref{eq:S-kinetic-term}) and $\braket{\Phi}$ (\ref{eq:Phi_kinetic_term}) gives 
\[ \label{eq:mass_terms}
& \frac{1}{2} \text{tr}\left( (D_\mu \braket{S})^\dagger  D^\mu \braket{S} \right)  + \frac{1}{2} \text{tr}\left( (D_\mu \braket{\Phi})^\dagger D^\mu \braket{\Phi} \right) \\
&= \frac{1}{2} g^2 v_{12}^2  w_1^2 
+ \frac{1}{2}g^2 ( v_{12}^2 + v_{23}^2 ) w_2^2 
+ \frac{1}{2}g^2 v_{23}^2 w_3^2 
+ \frac{1}{2}g^2 (v_S ^2 + v_{23}^2) w_{12}^2 \\
&+ \frac{1}{2}g^2 (4 v_S^2 + v_{12}^2 + v_{23}^2) w_{13}^2 
+ \frac{1}{2}g^2 (v_S^2 + v_{12}^2) w_{23}^2 
+ \frac{1}{2} g^2 \left( v_S^2 + 8v_{12}^2 + v_{23}^2  \right) z_{12}^2 \\
&+ \frac{1}{2}g^2 \left ( 4 v_S^2 + v_{12}^2 + v_{23}^2 \right)  z_{13}^2 
+ \frac{1}{2}g^2 \left( v_S^2 + v_{12}^2 + 8 v_{23}^2 \right)  z_{23}^2 
+ \frac{1}{2}g^2 ( - 2 v_{12}^2 )  w_1 w_2 \\
&+ \frac{1}{2}g^2( -2 v_{23}^2 ) w_2 w_3 
+ \frac{1}{2}g^2(2  v_{12} v_{23} ) w_{13} w_1 
+ \frac{1}{2}g^2(- 4 v_{12} v_{23} ) w_{13} w_2 \\
&+ \frac{1}{2}g^2(2 v_{12} v_{23} )  w_{13} w_3 
+ \frac{1}{2}g^2(- 2 v_{12} v_{23}) w_{12} w_{23}  
+ \frac{1}{2}g^2( -10 v_{12} v_{23} ) z_{12} z_{23}. \\
\]

\subsubsection{Mass eigenstate Gauge Fields}
\label{sec:mass_basis_gauge_fields}

In the previous section we found the gauge fields that acquired masses through the several stage breaking pattern. However, (\ref{eq:mass_terms}) is in the gauge basis and features several mixing terms between the states. To find the fields as mass eigenstates we write the Lagrangian as a quadratic form and diagonalise. For the sake of clarity we separate this into different quadratic forms. Firstly we have the mixing between $w_1,~ w_2, ~w_3$ and $w_{13}$ for which we can write 
\[
& \frac{1}{2} g^2 \begin{pmatrix}
     w_1 & w_2 & w_3 & w_{13} 
 \end{pmatrix} \begin{pmatrix}
     v_{12}^2 & -v_{12}^2 & 0 & v_{12} v_{23} \\ 
     -v_{12}^2 & v_{12}^2 + v_{23}^2 & -v_{23}^2 & -2v_{12} v_{23} \\ 
     0 & -v_{23}^2 & v_{23}^2 & v_{12} v_{23} \\ 
     v_{12} v_{23} & -2 v_{12} v_{23} & v_{12} v_{23} & v_{12}^2 + v_{23}^2 + 4 v_S^2 \\ 
 \end{pmatrix}
 \begin{pmatrix}
     w_1 \\ w_2 \\ w_3 \\ w_{13} 
 \end{pmatrix} \\
 =& \frac{1}{2} g^2 v_{12}^2 
 \begin{pmatrix}
     w_1 & w_2 & w_3 & w_{13} 
 \end{pmatrix} \begin{pmatrix}
     1 && -1 && 0 && \sqrt{\zeta} \\ 
     -1 && 1 + \zeta && -\zeta && -2\sqrt{\zeta} \\ 
     0 && -\zeta && \zeta && \sqrt{\zeta} \\ 
     \sqrt{\zeta} && -2 \sqrt{\zeta} && \sqrt{\zeta} && 1 + \zeta + 4 r \\ 
 \end{pmatrix}
 \begin{pmatrix}
     w_1 \\ w_2 \\ w_3 \\ w_{13} 
 \end{pmatrix},
\]
where we have defined $\zeta \equiv v_{23}^2/v_{12}^2 \ll 1$ and $r \equiv v_S^2/v_{12}^2 >1$. Determining the mass-eigenstates and corresponding eigenvalues we have to leading orders in $\zeta$ (noticing the prefactor of $g^2 v_{12}^2$)\footnote{Recall we have been suppressing SU(2) indices carried by the triplets.}
\[
W_{\text{SM}} = \frac{1}{\sqrt{3}} \sum_{i=1}^{3} w_i, \qquad & \qquad M^2_{W_{\text{SM}}}=0, \\
W_{23} \approx \frac{1}{\sqrt6} w_1 + \frac{1}{\sqrt6} w_2 - \sqrt{\frac{2}{3}}w_3, \qquad & \qquad M_{W_{23}}^2 \approx \frac{6r}{4r+1} g^2 v_{23}^2, \\
W_{12} \approx \frac{1}{\sqrt{2}} w_1 - \frac{1}{\sqrt{2}} w_2, \qquad & \qquad M^2_{W_{12}} \approx 2g^2 v_{12}^2, \\
\widetilde{W}_{13} \approx w_{13}, \qquad & \qquad M^2_{\widetilde{W}_{13}} \approx (4r+1)g^2 v_{12}^2.
\]
There is one eigenstate which is massless to all orders in $\zeta$, this is the SM $W$ triplet. However, the expressions for field redefinitions and mass eigenvalues are only approximate. In reality these are expressed as a series in  $\zeta$, so here we are only taking the leading order terms with the approximation approaching the true values when $v_{23}$ and $v_{12}$ becoming largely separated. We have also introduced a tilde notation – in the next section we will see that this identifies fields which have flavour off-diagonal couplings when coupling to fermions. This notation also identifies the fields which have masses $\sim v_{S}$, which arise from the initial $\Sp(6)_\text{L}$ breaking.

Next we deal with the mixing of the fields $w_{12}$ and $w_{23}$, where the relevant terms from (\ref{eq:mass_terms}) can be written as 
\[
& \frac{1}{2} g^2 \begin{pmatrix}
    w_{12} & w_{23}
\end{pmatrix}
\begin{pmatrix}
     v_S^2 + v_{23}^2 && - v_{12} v_{23} \\
    - v_{12} v_{23} && v_S^2 + v_{12}^2
\end{pmatrix}
\begin{pmatrix}
    w_{12} \\ w_{23}
\end{pmatrix} \\
& = \frac{1}{2} g^2 v_{12}^2 \begin{pmatrix}
    w_{12} & w_{23}
\end{pmatrix}
\begin{pmatrix}
    r+ \zeta  && - \sqrt{\zeta} \\
    - \sqrt{\zeta} && r+1
\end{pmatrix}
\begin{pmatrix}
    w_{12} \\ w_{23}
\end{pmatrix}.
\]
Diagonalising, we have the mass-eigenstates and eigenvalues
\[
\widetilde{W}_{12} \approx w_{12}, \qquad & \qquad M^2_{\widetilde{W}_{12}} = r g^2 v_{12}^2, \\
\widetilde{W}_{23} \approx w_{23}, \qquad & \qquad  M^2_{\widetilde{W}_{23}} = (r + 1) g^2 v_{12}^2.\\
\]

The final mixing is between the fields $z_{12}$ and $z_{23}$ which can be written as 
\[
& \frac{1}{2}g^2 \begin{pmatrix}
    z_{12} & z_{23} 
\end{pmatrix}
\begin{pmatrix}
    v_S^2 +8v_{12}^2 +v_{23}^2 && -5 v_{12} v_{23} \\
    -5 v_{12} v_{23} && v_S^2 + v_{12}^2 +8v_{23}^2
\end{pmatrix}
\begin{pmatrix}
    z_{12} \\
    z_{23}
\end{pmatrix} \\
& = \frac{1}{2}g^2 \begin{pmatrix}
    z_{12} & z_{23} 
\end{pmatrix}
\begin{pmatrix}
     r + 8 + \zeta && -5 \sqrt{\zeta} \\
    -5 \sqrt{\zeta} && r + 1 + 8 \zeta
\end{pmatrix}
\begin{pmatrix}
    z_{12} \\
    z_{23}
\end{pmatrix},
\]
Computing the mass-eigenvalues we find 
\[
Z_{23} \approx z_{23}, \qquad & \qquad M^2_{Z_{23}} = (r+1)g^2 v_{12}^2 \\
Z_{12} \approx  z_{12}, \qquad & \qquad M^2_{Z_{12}} = (r+8) g^2 v_{12}^2.
\]

The remaining field $z_{13}$ has no mixing and is already mass-diagonal, so trivially from (\ref{eq:mass_terms}) we read-off  

\[
Z_{13} = z_{13} \qquad \qquad M^2_{Z_{13}} \approx (4r+1)g^2v_{12}^2.
\]

Before considering interactions with fermions in the next section, we invert the definitions of the mass eigenstate fields to provide formulae of the gauge eigenstate fields in terms of the mass eigenstate fields:

\[
x^\mu_1 = \frac{1}{\sqrt{3}}W^\mu_{\text{SM}, I} \tau_I + \frac{1}{\sqrt{6}} W^\mu_{23,I} \tau_I + \frac{1}{\sqrt{2}} W_{12,I}^\mu \tau_I,
\]
\[
x^\mu_2 = \frac{1}{\sqrt{3}}W^\mu_{\text{SM},I} \tau_I + \frac{1}{\sqrt{6}} W_{23,I}^\mu \tau_I - \frac{1}{\sqrt{2}} W_{12,I}^\mu \tau_I,
\]
\[
x^\mu_3 = \frac{1}{\sqrt{3}}W_{\text{SM},I}^\mu \tau_I - \sqrt{\frac{2}{3}} W_{23,I}^\mu \tau_I ,
\]
\[
y^\mu_{12} = \widetilde{W}_{12,I}^\mu \tau_I + \im \frac{1}{2} Z^\mu_{12} \mathbb{I}_2,
\]
\[
y^\mu_{13} = \widetilde{W}_{13,I}^\mu \tau_I + \im \frac{1}{2} Z^\mu_{13} \mathbb{I}_2,
\]
\[
y^\mu_{23} = \widetilde{W}_{23,I}^\mu \tau_I + \im \frac{1}{2} Z^\mu_{23} \mathbb{I}_2.
\]
Here we are using $x_i $ and $y_{ij}$ which reintroduces the matrix structure, this will be vital in the next section for determining the couplings to SM fields.

\subsection{Gauge-fermion interactions} \label{sec:gauge-fermion}

Writing the gauge potential $\mathcal{W}^\mu_AT_A$ in terms of $x^\mu_i$ and $y^\mu_{ij}$ via (\ref{eq:sp(6)elem}), the gauge-fermion interactions for the LH quarks are given by
\[
\mathcal{L}_{\text{quark-gauge}} &= g \left( \bar{Q}' \mathcal{W}^\mu_A T_A \gamma^\mu Q'\right) \\
&= g \sum_{i=1}^3 \left(\bar{q}'_i x_{i}^\mu \gamma_\mu q'_i  \right) + g \sum_{i < j} \left(\bar{q}'_i y_{ij}^\mu \gamma_\mu q'_j  \right) + g \sum_{i > j} \left(\bar{q}'_i y^{\mu \dagger}_{ji} \gamma_\mu q'_j  \right),
\]
where $g$ is the $\Sp(6)_\text{L}$ coupling and the prime ($'$) denoting that the quarks fields are in the gauge basis and are have not yet been rotated to the mass basis. In terms of the mass eigenstate fields we have the
\[ \label{eq:quark-gauge}
\mathcal{L}_{\text{quark-gauge}} &= \frac{1}{\sqrt{3}} g W^{\mu I}_\text{SM} \sum_i  (\bar{q}'_i \gamma_\mu \tau^I q'_i) \\
&+ g W_{12}^{\mu I} \left( \frac{1}{\sqrt{2}} \left( \bar{q}'_1 \gamma_\mu \tau^I q'_1 \right) - \frac{1}{\sqrt{2}}\left( \bar{q}'_2 \gamma_\mu \tau^I q'_2 \right) \right) \\
&+ g W_{23}^{\mu I} \left( \frac{1}{\sqrt{6}} \left( \bar{q}'_1 \gamma_\mu \tau^I q'_1 \right) + \frac{1}{\sqrt{6}} \left( \bar{q}'_2 \gamma_\mu \tau^I q'_2 \right) - \sqrt{\frac{2}{3}} \left( \bar{q}'_3 \gamma_\mu \tau^I q'_3 \right)  \right) \\
&+ g \widetilde{W}_{13}^{\mu I} \left( \left( \bar{q}'_1 \gamma_\mu \tau^I q'_3 \right) + \left( \bar{q}'_3 \gamma_\mu \tau^I q'_1 \right)\right) \\
&+ g \widetilde{W}_{12}^{\mu I} \left( \left( \bar{q}'_1 \gamma_\mu \tau^I q'_2 \right) + \left( \bar{q}'_2 \gamma_\mu \tau^I q'_1 \right)\right) \\
&+ g \widetilde{W}_{23}^{\mu I} \left( \left( \bar{q}'_2 \gamma_\mu \tau^I q'_3 \right) + \left( \bar{q}'_3 \gamma_\mu \tau^I q'_2 \right)\right) \\
&+ \frac{g}{2} Z_{12}^\mu \left( \im\left( \bar{q}'_1 \gamma_\mu \mathbb{I}_2 q'_2  \right) - \im \left( \bar{q}'_2 \gamma_\mu \mathbb{I}_2 q'_1  \right)  \right) \\
&+ \frac{g}{2} Z_{13}^\mu \left( \im\left( \bar{q}'_1 \gamma_\mu \mathbb{I}_2 q'_3  \right) - \im \left( \bar{q}'_3 \gamma_\mu \mathbb{I}_2 q'_1  \right)  \right) \\
&+ \frac{g}{2} Z_{23}^\mu \left( \im\left( \bar{q}'_2 \gamma_\mu \mathbb{I}_2 q'_3  \right) - \im \left( \bar{q}'_3 \gamma_\mu \mathbb{I}_2 q'_2  \right)  \right).
\]
Examining the term involving $W_\text{SM}$, it is apparent that the $\Sp(6)_\text{L}$ coupling, $g$, is related to the SM $\SU(2)_\text{L}$ coupling by
\[ \label{eq:coupling_relation}
g_{\text{SM}} = \frac{1}{\sqrt{3}}g.
\]

Since (\ref{eq:quark-gauge}) is still in terms of gauge-basis quark fields, it remains to rotate them to the mass basis fields. These are related by 
\[
u_i = (V_u)_{ij} u'_j, \qquad d_i = (V_d)_{ij} d'_j,
\] 
such that for the quark doublets of the SM we can write
\[ \label{eq:Vu_def}
q'_i = \begin{pmatrix}
    u'_i \\ d'_i
\end{pmatrix} = \begin{pmatrix}
    (V^\dagger_u)_{ij} u_j \\ (V^\dagger_d)_{ij} d_j
\end{pmatrix} =
(V^\dagger_u)_{ij}
\begin{pmatrix}
    u_j \\ V_{jk} d_k
\end{pmatrix} =
(V^\dagger_u)_{ij} q_j,
\]
where $V \equiv V_u V^\dagger_d$ is the CKM matrix and $q_i \equiv (u_i, V_{ij} d_j)^\text{T}$. By defining flavour coupling matrices $[g]_{ij}$ for each gauge field, we can write the Lagrangian as (with implied sum over repeated flavour indices)
\[
\mathcal{L}_\text{quark-gauge} &= g_\text{SM} W^{\mu I}_\text{SM}  (\bar{q}_i \gamma_\mu \tau^I q_i) + \left[ g^q_{W_{12}} \right]_{ij} W_{12}^{\mu I} (\bar{q}_i \gamma_\mu \tau^I q_j) + \left[ g^q_{W_{23}} \right]_{ij} W_{23}^{\mu I} (\bar{q}_i \gamma_\mu \tau^I q_j) \\
&+ \left[ g^q_{\widetilde{W}_{13}} \right]_{ij} \widetilde{W}_{13}^{\mu I} (\bar{q}_i \gamma_\mu \tau^I q_j) + \left[ g^q_{\widetilde{W}_{12}} \right]_{ij} \widetilde{W}_{12}^{\mu I} (\bar{q}_i \gamma_\mu \tau^I q_j) + \left[ g^q_{\widetilde{W}_{23}} \right]_{ij} \widetilde{W}_{23}^{\mu I} (\bar{q}_i \gamma_\mu \tau^I q_j) \\
&+ \left[ g^q_{Z_{12}} \right]_{ij} Z_{12}^{\mu} (\bar{q}_i \gamma_\mu  q_j) + \left[ g^q_{Z_{13}} \right]_{ij} Z_{13}^{\mu} (\bar{q}_i \gamma_\mu  q_j) + \left[ g^q_{Z_{23}} \right]_{ij} Z_{23}^{\mu} (\bar{q}_i \gamma_\mu q_j)
\]
where the hermitian coupling matrices\footnote{In the formulation of the $\SU(2)^3_\text{L}$ model \cite{Davighi:2023xqn}, a parametrization was introduced through spherical polar coordinates $\theta$ and $\phi$. This model imposes the angular variables to take the values $\phi= \pi/4$ and $\theta = \tan^{-1}(\sqrt{2})$.} are given by 
\[ \label{eq:q_couplings}
\left[ g^q_{W_{12}} \right] &= \sqrt{\frac{3}{2}} g_\text{SM} V_u \begin{pmatrix}
    1 & 0 & 0 \\
    0 & -1 & 0  \\
    0 & 0 & 0 
\end{pmatrix} V^\dagger_u, \quad 
\left[ g^q_{W_{23}} \right] = \frac{1}{\sqrt{2}} g_\text{SM} V_u \begin{pmatrix}
    1 & 0 & 0 \\
    0 & 1 & 0  \\
    0 & 0 & -2 
\end{pmatrix} V^\dagger_u, \\
\left[ g^q_{\widetilde{W}_{13}} \right] &=  \sqrt{3} g_\text{SM} V_u \begin{pmatrix}
    0 & 0 & 1 \\
    0 & 0 & 0  \\
    1 & 0 & 0 
\end{pmatrix} V^\dagger_u, \quad 
\left[ g^q_{\widetilde{W}_{12}} \right] =  \sqrt{3} g_\text{SM} V_u \begin{pmatrix}
    0 & 1 & 0 \\
    1 & 0 & 0  \\
    0 & 0 & 0 
\end{pmatrix} V^\dagger_u, \\
\left[ g^q_{\widetilde{W}_{23}} \right] &=  \sqrt{3} g_\text{SM} V_u \begin{pmatrix}
    0 & 0 & 0 \\
    0 & 0 & 1  \\
    0 & 1 & 0 
\end{pmatrix} V^\dagger_u, \quad 
\left[ g^q_{Z_{12}} \right] =  \frac{\sqrt{3}}{2} g_\text{SM} V_u \begin{pmatrix}
    0 & \im & 0 \\
    -\im & 0 & 0  \\
    0 & 0 & 0 
\end{pmatrix} V^\dagger_u, \\
\left[ g^q_{Z_{13}} \right] &=  \frac{\sqrt{3}}{2} g_\text{SM} V_u \begin{pmatrix}
    0 & 0 & \im \\
    0 & 0 & 0  \\
    -\im & 0 & 0 
\end{pmatrix} V^\dagger_u, \quad
\left[ g^q_{Z_{23}} \right] =  \frac{\sqrt{3}}{2} g_\text{SM} V_u \begin{pmatrix}
    0 & 0 & 0 \\
    0 & 0 & \im  \\
    0 & -\im & 0 
\end{pmatrix} V^\dagger_u.
\]
It is worth noting that the couplings here depend on the relative alignment between the quark mass basis and the gauge basis.\footnote{There is nothing special about the $u$ quarks here: equivalently, one may choose a basis which prioritises the down-type quarks.} With the convention $V=V_u V_d^\dagger$, a choice of $V_u$ fixes $V_d=V^\dagger V_u$. Two particularly simple benchmark choices are $V_u=\mathbb{I}$, for which $V_d=V^\dagger$, which we refer to as \textit{up-aligned}, and $V_u=V$, for which $V_d=\mathbb{I}$, which we refer to as \textit{down-aligned}. These two cases should not be interpreted as providing rigorous upper and lower bounds for an arbitrary unitary $V_u$. In particular, a generic $V_u$ may contain additional phases which can affect CP-violating meson-mixing observables. The limits derived below should therefore be understood as constraints in these two benchmark alignment scenarios, rather than as an exhaustive scan over the full flavour-alignment parameter space.

We should also consider the interactions between the leptons and the gauge fields. These are analogous to the quark terms, given by
\[
\mathcal{L}_\text{lepton-gauge} &= g_\text{SM} W^{\mu I}_\text{SM}  (\bar{l}_i \gamma_\mu \tau^I l_i) + \left[ g^l_{W_{12}} \right]_{ij} W_{12}^{\mu I} (\bar{l}_i \gamma_\mu \tau^I l_j) + \left[ g^l_{W_{23}} \right]_{ij} W_{23}^{\mu I} (\bar{l}_i \gamma_\mu \tau^I l_j) \\
&+ \left[ g^l_{\widetilde{W}_{13}} \right]_{ij} \widetilde{W}_{13}^{\mu I} (\bar{l}_i \gamma_\mu \tau^I l_j) + \left[ g^l_{\widetilde{W}_{12}} \right]_{ij} \widetilde{W}_{12}^{\mu I} (\bar{l}_i \gamma_\mu \tau^I l_j) + \left[ g^l_{\widetilde{W}_{23}} \right]_{ij} \widetilde{W}_{23}^{\mu I} (\bar{l}_i \gamma_\mu \tau^I l_j) \\
&+ \left[ g^l_{Z_{12}} \right]_{ij} Z_{12}^{\mu} (\bar{l}_i \gamma_\mu  l_j) + \left[ g^l_{Z_{13}} \right]_{ij} Z_{13}^{\mu} (\bar{l}_i \gamma_\mu  l_j) + \left[ g^l_{Z_{23}} \right]_{ij} Z_{23}^{\mu} (\bar{l}_i \gamma_\mu l_j)
\]
with coupling matrices given by 
\[ \label{eq:l_couplings}
\left[ g^l_{W_{12}} \right] &= \sqrt{\frac{3}{2}} g_\text{SM} V_l \begin{pmatrix}
    1 & 0 & 0 \\
    0 & -1 & 0  \\
    0 & 0 & 0 
\end{pmatrix} V^\dagger_l, \quad 
\left[ g^l_{W_{23}} \right] = \frac{1}{\sqrt{2}} g_\text{SM} V_l \begin{pmatrix}
    1 & 0 & 0 \\
    0 & 1 & 0  \\
    0 & 0 & -2 
\end{pmatrix} V^\dagger_l, \\
\left[ g^l_{\widetilde{W}_{13}} \right] &=  \sqrt{3} g_\text{SM} V_l \begin{pmatrix}
    0 & 0 & 1 \\
    0 & 0 & 0  \\
    1 & 0 & 0 
\end{pmatrix} V^\dagger_l, \quad 
\left[ g^l_{\widetilde{W}_{12}} \right] =  \sqrt{3} g_\text{SM} V_l \begin{pmatrix}
    0 & 1 & 0 \\
    1 & 0 & 0  \\
    0 & 0 & 0 
\end{pmatrix} V^\dagger_l, \\
\left[ g^l_{\widetilde{W}_{23}} \right] &=  \sqrt{3} g_\text{SM} V_l \begin{pmatrix}
    0 & 0 & 0 \\
    0 & 0 & 1  \\
    0 & 1 & 0 
\end{pmatrix} V^\dagger_l, \quad 
\left[ g^l_{Z_{12}} \right] =  \frac{\sqrt{3}}{2} g_\text{SM} V_l \begin{pmatrix}
    0 & \im & 0 \\
    -\im & 0 & 0  \\
    0 & 0 & 0 
\end{pmatrix} V^\dagger_l, \\
\left[ g^l_{Z_{13}} \right] &=  \frac{\sqrt{3}}{2} g_\text{SM} V_l \begin{pmatrix}
    0 & 0 & \im \\
    0 & 0 & 0  \\
    -\im & 0 & 0 
\end{pmatrix} V^\dagger_l, \quad
\left[ g^l_{Z_{23}} \right] =  \frac{\sqrt{3}}{2} g_\text{SM} V_l \begin{pmatrix}
    0 & 0 & 0 \\
    0 & 0 & \im  \\
    0 & -\im & 0 
\end{pmatrix} V^\dagger_l.
\]
Since $V_l$ is unphysical in the SM $V_l$ can be an arbitrary unitary matrix. If we ignore CP violation effects in the charged leptons, we can approximate this matrix as an orthogonal matrix
\[ \label{eq:Vl}
V_l = \begin{pmatrix}
    \cos \theta_{12} & -\sin\theta_{12} & 0 \\
    \sin \theta_{12} & \cos \theta_{12} & 0 \\
    0 & 0 & 1 \\
\end{pmatrix}
\begin{pmatrix}
    \cos \theta_{13} & 0 & \sin \theta_{13} \\
    0 & 1 & 0 \\
    -\sin \theta_{13} & 0 & \cos \theta_{13} \\
\end{pmatrix}
\begin{pmatrix}
    1 & 0 & 0 \\
    0 & \cos \theta_{23} & -\sin \theta_{23} \\
    0 & \sin \theta_{23}  & \cos \theta_{23} \\
\end{pmatrix}.
\]
Further if we take these angles to be small we have the approximation
\[
V_l \approx \left(
\begin{array}{ccc}
 1-\frac{\theta_{12} ^2}{2}-\frac{\theta_{13} ^2}{2} &  -\theta_{12} +\theta_{13}  \theta_{23}  & \theta_{13} +\theta_{12}  \theta_{23}   \\
 \theta_{12}  & 1-\frac{\theta_{12} ^2}{2}-\frac{\theta_{23} ^2}{2} & -\theta_{23}+ \theta_{12}  \theta_{13}   \\
 -\theta_{13}  & \theta_{23}  & 1-\frac{\theta_{13} ^2}{2}-\frac{\theta_{23} ^2}{2} \\
\end{array}
\right)\]

Observables involving leptons will of course depend on these angles, which can be arbitrarily chosen. We consider two benchmarks corresponding to different hierarchies in these angles, these consist of a CKM-like hierarchy
\[ \label{eq:Vl_1}
\theta_{13} \ll \theta_{23} \ll \theta_{12}: \quad \theta_{12} = \alpha, \quad \theta_{23} = \alpha^2, \quad \theta_{13} = \alpha^3,
\]
and 
\[ \label{eq:Vl_2}
\theta_{13} \ll \theta_{12}=\theta_{23}: \quad\theta_{12} = \alpha, \quad \theta_{23} = \alpha, \quad \theta_{13} = \alpha^2.
\]
Since $V_l$ is unphysical in the SM and is not predicted by the minimal model considered here, the lepton-flavour observables discussed below should be interpreted as benchmark-dependent probes rather than sharp model predictions. In particular, the rates for $\mu\to 3e$ and $\mu\to e$ conversion which we examine in §\ref{sec:pheno} depend on the charged-lepton mass--gauge misalignment encoded in $V_l$. We therefore use the hierarchies in (\ref{eq:Vl_1}) and (\ref{eq:Vl_2}) as representative choices, rather than as predictions of the UV theory.

\subsection{Gauge-Higgs interactions}

As stated in §\ref{sec:Group&Fields} this model features three flavours of Higgs doublets. The gauge fields will therefore mediate interactions between the SM Higgs $H_3$ and the BSM Higgs doublets $H_1$ and $H_2$. 

In the UV, the Higgs multiplet $\mathcal{H}$ has covariant derivative given by
\[
D^\mu \mathcal{H} = \partial^\mu \mathcal{H} - \im g \mathcal{W}_A^\mu T_A \mathcal{H}.
\]
As before, we can write first $\mathcal{W}^\mu_A T_A$ in terms of the $x_i^\mu$ and $y_{ij}^\mu$ as in (\ref{eq:sp(6)elem}) and write $\mathcal{H} = (H_1, H_2, H_3)^\text{T}$, where $H_3$ is the SM Higgs and $H_1, ~ H_2$ are assumed to be heavy. Considering the $i^\text{th}$ doublet of $\mathcal{H}$, we have 
\[
D_\mu H_i =  \partial_\mu H_i - \im g \sum_{j = i} x_i H_j - \im g \sum_{j > i} y_{i j} H_j - \im g \sum_{j < i } y^\dagger_{j i} H_j. 
\]
In terms of the mass eigenstate fields, the covariant derivatives for the three doublets are 
\[
D^\mu H_1 &= \partial^\mu H_1 - \im g  x_1^\mu H_1 -\im g y_{12}^\mu H_2 -\im g y^\mu_{13} H_3 \\
&=  \partial^\mu H_1 - \im g \left(  \frac{1}{\sqrt{3}}W^\mu_{\text{SM}, I} \tau_I + \frac{1}{\sqrt{6}} W^\mu_{23,I} \tau_I + \frac{1}{\sqrt{2}} W_{12,I}^\mu \tau_I \right) H_1 \\ 
&-\im g \left( \widetilde{W}_{12,I}^\mu \tau_I + \im \frac{1}{2} Z^\mu_{12} \mathbb{I}_2 \right) H_2 -\im g \left( \widetilde{W}_{13,I}^\mu \tau_I + \im \frac{1}{2} Z^\mu_{13,I} \mathbb{I}_2 \right) H_3 
\]
\[
D^\mu H_2 &= \partial^\mu H_2 - \im g x_2^\mu H_2 - \im g y_{12}^{\mu \dagger} H_1 - \im g y_{23} H_3 \\
&= \partial^\mu H_2 - \im g \left(  \frac{1}{\sqrt{3}}W^\mu_{\text{SM},I} \tau_I + \frac{1}{\sqrt{6}} W_{23,I}^\mu \tau_I - \frac{1}{\sqrt{2}} W_{12,I}^\mu \tau_I \right) H_2 \\
&-\im g \left( \widetilde{W}_{12,I}^\mu \tau_I - \im \frac{1}{2} Z^\mu_{12} \mathbb{I}_2 \right) H_1 -\im g \left( \widetilde{W}_{23,I}^\mu \tau_I + \im \frac{1}{2} Z^\mu_{23} \mathbb{I}_2 \right) H_3 \\
\]
\[
D^\mu H_3 &= \partial^\mu H_3 - \im g x_3^\mu H_3 -\im g y_{13}^{\mu\dagger} H_1 -\im g y^{\mu\dagger}_{23} H_2  \\
 &= \partial^\mu H_3 - \im g \left( \frac{1}{\sqrt{3}}W_{\text{SM},I}^\mu \tau_I - \sqrt{\frac{2}{3}} W_{23,I}^\mu \tau_I \right) H_3 \\
 &-\im g \left( \widetilde{W}_{13,I}^\mu \tau_I - \im \frac{1}{2} Z^\mu_{13} \mathbb{I}_2 \right) H_1 -\im g \left( \widetilde{W}_{23,I}^\mu \tau_I - \im \frac{1}{2} Z^\mu_{23} \mathbb{I}_2 \right) H_2  \\
\] 
With these covariant derivatives, we can compute the UV kinetic term 
\[
|D_\mu \mathcal{H}|^2 = |D_\mu H_1|^2 + |D_\mu H_2|^2 + |D_\mu H_3|^2,
\]
which encodes all the gauge-Higgs interactions. For simplicity, we only consider the interactions coming from $|D_\mu H_3|^2$ which is the most important since it contains the interactions involving the SM Higgs. In particular, this term contains the interaction
\[
|D_\mu H_3|^2 \supset -\sqrt{2} g_\text{SM} W^{\mu}_{23,I} H_3^\dagger \tau_I \im D_{\text{SM} \mu} H_3 + \text{h.c.},
\]
where $D^\mu_\text{SM} H_3 = \partial^\mu H_3 - \im g_\text{SM} W^\mu_{SM,I} \tau_I H_3 $ is the SM $\SU(2)_\text{L}$ covariant derivative. 

This term is particularly important since it can modify couplings to the SM $W$ and $Z$ bosons, and additionally generates dimension-6 SMEFT operators as we shall see in the next section. For this purpose, in a similar manner to the quark and lepton interactions we write this interaction as 
\[
|D_\mu H_3|^2 \supset \left(g^H_{{W_{23}}} \right) W^{\mu}_{23,I} H_3^\dagger \tau_I \im D_{\text{SM} \mu} H_3 + \text{h.c.},
\]
with 
\[
g^H_{{W_{23}}} = - \sqrt{2} g_\text{SM}.
\]

\subsection{Comments on the Scalar Sector and Yukawa Completion}
\label{sec:scalar_yukawa_completion}

We now return to the scalar and Yukawa issues mentioned above. The minimal renormalisable Yukawa sector, together with the assumption that only one Higgs doublet acquires a vev, gives rank-one Yukawa matrices. Moreover, the vacuum alignment required to keep the remaining doublets inert is not fixed by the gauge-sector analysis alone.

Before considering these issues, it is worth mentioning why the phenomenological focus on the gauge sector over the scalar sector in this model is well motivated. The main reason is that $S$ and $\Phi$ cannot form renormalisable Yukawa interactions with the fermions. Since both representations carry two $\Sp(6)_\text{L}$ indices, forming a Yukawa-type gauge invariant operator would require a left-handed fermion transforming in the fundamental representation of $\Sp(6)_\text{L}$, but also a right-handed fermion transforming in the same representation. Since the SM fermions are singlets under $\Sp(6)_\text{L}$, no such operators exist. Therefore the only particles of the SM that $S$ and $\Phi$ couple to are the SM Higgs and the SM $W$ triplet, though since the most constraining observables come from flavour observations of fermions, the BSM scalar effects are assumed to be subleading to the gauge BSM effects.

In §\ref{sec:Group&Fields} it was hypothesised that, of the three electroweak doublets emerging from $\mathcal{H}$, only one linear combination obtains a vev. This can be understood directly from the scalar potential in (\ref{eq:higgs_potential}). Under the decomposition $\mathcal{H}=\bigoplus_{i=1}^3 H_i$, the relevant terms take the form
\[
V
=
\sum_{i,j=1}^3
H_i^\dagger M_{ij} H_j
+
\lambda
\left(
\sum_{i=1}^3 H_i^\dagger H_i
\right)^2 .
\]
The mass matrix $M$ is generated by the dimension-three and dimension-four interactions involving $\mathcal{H}$, $S$ and $\Phi$ after insertion of the symmetry-breaking vevs\footnote{The unusual cubic terms are allowed since the fundamental representation of $\Sp(6)_\text{L}$ obeys $\vec{6}\otimes \bar{\vec{6}}=\vec{21}\oplus \vec{14}\oplus \vec{1}$.}. In general $M$ is diagonalised by a unitary transformation $H_i=U_{ij}H_j'$, while the quartic term is invariant under this rotation. The scalar potential may therefore be written as
\[
V
=
\sum_{i=1}^3
m_i^2 H_i^{\prime\dagger}H_i'
+
\lambda
\left(
\sum_{i=1}^3 H_i^{\prime\dagger}H_i'
\right)^2 .
\]
For $\lambda>0$, the vacuum is aligned with the doublet whose mass parameter is most negative. Indeed, if two non-degenerate doublets were to acquire vevs, the corresponding minimisation conditions would require the same value of $m_i^2$ for both fields. Thus, away from degenerate special points, the generic vacuum contains a vev for only a single Higgs doublet. We identify this light direction with $H_3$, with $\braket{H_3}=(0,v_H)^T$, while the orthogonal doublets remain inert heavy states. A more detailed study of the multi-Higgs potential implied by the $\Sp(6)_\text{L}$ structure will be presented in future work.

With a vev aligned with a single doublet of $\H$, from (\ref{eq:yukawa_matrices}) it is clear that the Mass matrices formed from the Higgs Yukawa couplings (\ref{eq:UV_yukawas}) are rank-1, which have a single mass eigenvalue. In particular the mass matrices can be written in the form of an outer product
\[
M_\psi = v_H \begin{pmatrix}
    0 \\ 0 \\ 1
\end{pmatrix} \begin{pmatrix}
    y_{\psi,1} & y_{\psi_2} & y_{\psi,3}
\end{pmatrix}.
\]

A common approach to introduce the additional degrees of freedom required in $M_\psi$ to lift the rank, is to form EFT operators with the link fields of the model. For example Ref. \cite{Davighi:2023xqn}, utilises $\phi_{12}$ and $\phi_{23}$, which are common to this model, to form EFT operators which generates a further two mass eigenvalues which are hierarchical. These EFT operators can be formed by integrating out the heavy Higgses $H_1$ or $H_2$, but due to the symmetries of this model such operators do not lift the rank of the Yukawas. In particular, such EFT operators are not independent of the Yukawas (\ref{eq:UV_yukawas}).

As an example, consider the UV cubic interaction which generates the Higgs mixing term
\[
\mathcal{L} \supset  \mu_{\H \Phi} \H^\dagger \Phi \Omega \H \supset -\mu_{\H \Phi}  \left( v_{12} H_1^\dagger H_2 + v_{23} H_{2}^\dagger H_3\right),
\]
noting that $\braket{\phi_{ij}} \epsilon = -v_{ij } \mathbb{I}_2$. From the Yukawa terms (\ref{eq:intermediate_yukawas}), the dependence on the left-handed family index is entirely fixed by the Higgs field which appears in the operator. Taking $H_1$ and $H_2$ to be heavy, with masses $M_1$ and $M_2$, they can be integrated out giving
\[
H_2
\simeq
-\frac{\mu_{\mathcal{H}\Phi}v_{23}}{M_2^2}H_3.
\qquad 
H_1
\simeq
\frac{\mu_{\mathcal{H}\Phi}^2v_{12}v_{23}}{M_1^2M_2^2}H_3.
\]
which generate the effective Yukawa interactions
\[
\mathcal{L}_{\rm EFT}
& \supset 
\sum_j
y_{e,j}\,\bar{l}_3H_3e_j + y_{d,j}\,\bar{q}_3H_3d_j + y_{u,j}\,\bar{q}_3\widetilde{H}_3u_j \\
&-
\sum_j
\frac{\mu_{\mathcal{H}\Phi}v_{23}}{M_2^2}
\left(
y_{e,j}\,\bar{l}_2H_3e_j + y_{d,j}\,\bar{q}_2H_3d_j
\right)
-
\sum_j
\frac{\mu_{\mathcal{H}\Phi}^\ast v_{23}^\ast}{M_2^2}
y_{u,j}\,\bar{q}_2\widetilde{H}_3u_j \\
&+
\sum_j
\frac{\mu_{\mathcal{H}\Phi}^2v_{12}v_{23}}{M_1^2M_2^2}
\left(
y_{e,j}\,\bar{l}_1H_3e_j + y_{d,j}\,\bar{q}_1H_3d_j \right)
+
\sum_j \frac{(\mu^*_{\mathcal{H}\Phi})^2v_{12} v_{23}}{M_1^2M_2^2} y_{u,j}\,\bar{q}_1\widetilde{H}_3u_j
+
\text{h.c.}
\]
Therefore, the induced Yukawa entries for the first and second left-handed families are generated by the Higgs mixing terms. However, \textit{they are not independent of the original UV Yukawa couplings}. Defining
\[
\zeta_{23}
=
-\frac{\mu_{\mathcal{H}\Phi}v_{23}}{M_2^2},
\qquad
\zeta_{12}
=
-\frac{\mu_{\mathcal{H}\Phi}v_{12}}{M_1^2},
\]
the mass matrices take the schematic form
\[
M_{\psi}^{\rm eff}
= v
\begin{pmatrix}
\zeta_{12}\zeta_{23} \\
\zeta_{23} \\
1
\end{pmatrix}
\begin{pmatrix}
y_{\psi,1} & y_{\psi,2} & y_{\psi,3}
\end{pmatrix},
\]

The precise signs in $\zeta_{12}$ and $\zeta_{23}$ follow from the convention chosen for the mixed mass terms in the Lagrangian. The crucial point is that the effective Yukawa matrix is still an outer product of a single vector in left-handed flavour space with a single vector in right-handed flavour space. The EFT operators generated by integrating out $H_1$ and $H_2$ therefore populate the first and second rows of the Yukawa matrices, but they do so in a way which is proportional to the third row. They generate a hierarchy among the rows, but not new independent flavour directions. As a result, the SM fermion mass matrices remain rank-1 and only a single fermion obtains a mass in each sector.

This should be understood as a limitation of the minimal implementation considered here, rather than as an obstruction to the broader class of UV completions. In this model, flavour is only embedded in the left-handed gauge structure, through $\Sp(6)_\text{L}$. The right-handed SM fermions are not unified in an analogous flavour gauge structure, and hence the Higgs mixing effects above can only generate structure in the left-handed family index. Since the right-handed index is still controlled by the single UV Yukawa vector $y_{\psi,j}$, the resulting Yukawa matrices remain rank-1.

A more complete theory of flavour would therefore require additional ingredients which generate independent structures in the right-handed flavour space. One natural possibility is that the present setup is embedded into a larger framework, for example of the form $\SU(4)\times \Sp(6)_\text{L}\times \Sp(6)_\text{R}$ \cite{Davighi:2022fer}, in which the right-handed fermions are also charged under a non-trivial flavour gauge symmetry. Alternatively, extra vector-like fermions in the UV could mediate additional gauge-invariant interactions, leading after their decoupling to further effective Yukawa operators with independent flavour spurions. In such extensions the EFT operators need not be proportional to the same vector $y_{\psi,j}$, and the rank of the SM Yukawa matrices may be lifted. The rank-1 result found above is therefore a consequence of the minimal left-handed flavour embedding, rather than a generic prediction of all possible UV completions.

\subsection{Summary}

\begin{table}[h]
    \centering
    \begin{tabular}{|c|c|c|c|c|}
    \hline
        Field Name & SM Irrep & $M^2$ & SM quark/fermion coupling & SM Higgs coupling \\
        \hline
        & & & & \\
        $W_{23}$ & $(\vec{1},\vec{3})_0$ & $\frac{18r}{4r+1} g_\text{SM}^2 v_{23}^2$ & $\frac{1}{\sqrt{2}} g_\text{SM} V_{u(l)} \begin{pmatrix}
            1 & 0 & 0 \\ 0 & 1 & 0 \\ 0 & 0 & -2 
        \end{pmatrix} V_{u(l)}^\dagger$ & $-\sqrt{2} g_\text{SM}$ \\
        & & & & \\
        $W_{12}$ & $(\vec{1},\vec{3})_0$ & $6 g_\text{SM}^2 v_{12}^2$ & $\sqrt{\frac{3}{2}} g_\text{SM} V_{u(l)} \begin{pmatrix}
            1 & 0 & 0 \\ 0 & -1 & 0 \\ 0 & 0 & 0 
        \end{pmatrix} V_{u(l)}^\dagger$ & \\
        & & & & \\
        $\widetilde{W}_{12}$ & $(\vec{1},\vec{3})_0$ & $3r g_\text{SM}^2 v_{12}^2$ &  $\sqrt{3} g_\text{SM} V_{u(l)} \begin{pmatrix}
            0 & 1 & 0 \\ 1 & 0 & 0 \\ 0 & 0 & 0 
        \end{pmatrix} V_{u(l)}^\dagger$ & \\
        & & & & \\
        $\widetilde{W}_{13}$ & $(\vec{1},\vec{3})_0$ & $3(4r+1)g_\text{SM}^2 v_{12}^2$  & $\sqrt{3} g_\text{SM} V_{u(l)} \begin{pmatrix}
            0 & 0 & 1 \\ 0 & 0 & 0 \\ 1 & 0 & 0 
        \end{pmatrix} V_{u(l)}^\dagger$ & \\
        & & & & \\
        $\widetilde{W}_{23}$ & $(\vec{1},\vec{3})_0$ & $3(r+1) g_\text{SM}^2 v_{12}^2$  & $\sqrt{3} g_\text{SM} V_{u(l)} \begin{pmatrix}
            0 & 0 & 0 \\ 0 & 0 & 1 \\ 0 & 1 & 0 
        \end{pmatrix} V_{u(l)}^\dagger$ & \\
        & & & & \\
        $Z_{12}$ & $(\vec{1},\vec{1})_0$ & $3(r+8)g_\text{SM}^2 v_{12}^2$ & $\frac{\sqrt3}{2} g_\text{SM} V_{u(l)} \begin{pmatrix}
            0 & \im & 0 \\ -\im & 0 & 0 \\ 0 & 0 & 0 
        \end{pmatrix} V_{u(l)}^\dagger$ & \\
        & & & & \\
        $Z_{13}$ & $(\vec{1},\vec{1})_0$ & $3(4r+1) g_\text{SM}^2 v_{12}^2$ & $\frac{\sqrt3}{2} g_\text{SM} V_{u(l)} \begin{pmatrix}
            0 & 0 & \im \\ 0 & 0 & 0 \\ -\im & 0 & 0 
        \end{pmatrix} V_{u(l)}^\dagger$ & \\
        & & & & \\
        $Z_{23}$ & $(\vec{1},\vec{1})_0$ & $3(r+1) g_\text{SM}^2 v_{12}^2$ & $\frac{\sqrt3}{2} g_\text{SM} V_{u(l)} \begin{pmatrix}
            0 & 0 & 0 \\ 0 & 0 & \im \\ 0 & -\im & 0 
        \end{pmatrix} V_{u(l)}^\dagger$  & \\
        & & & & \\
        \hline
    \end{tabular}
    \caption{Summary table of the BSM gauge fields of this model along with their SM representations, tree-level masses (squared), flavour coupling matrices to SM quark and lepton doublets, and their couplings to the SM Higgs.}
    \label{tab:GuageBosons}
\end{table}

To summarise, the breaking at three distinct scales ($v_{23}, ~v_{12}$ and $v_S$) introduces eight BSM gauge fields. Table \ref{tab:GuageBosons} gives a full summary of the tree-level masses of these gauge fields and couplings to the SM fermions and Higgs, with these masses differing from those derived in §\ref{sec:mass_basis_gauge_fields} since they are written in terms of $g_\text{SM}$.

At the highest scale $v_S$, as representations of the SM $\SU(2)_\text{L}$ we have three triplets $\widetilde{W}_{12}$, $\widetilde{W}_{13}$ and  $\widetilde{W}_{23}$, in addition to three singlets $Z_{12}$, $Z_{13}$ and $Z_{23}$ which result from breaking $\Sp(6)_\text{L} \rightarrow \SU(2)^3_\text{L}$. These fields feature off-diagonal flavour couplings to the gauge basis SM fermions and can mediate flavour changing processes even in the case of alignment between the mass and gauge bases of the fermions.

Below this scale the model is similar to the $\SU(2)^3_\text{L}$ model \cite{Davighi:2023xqn}, with another two triplets $W_{12}$ and $W_{23}$ at the scales $v_{12}$ and $v_{23}$, respectively. These fields feature diagonal but \text{not} flavour universal couplings to the gauge basis SM fermions, with the $W_{23}$ allowed to be fairly light by coupling flavour universally to the first two generations of fermions. Additionally, the $W_{23}$ is the only BSM field to have consequential couplings to the SM Higgs. 

\begin{figure}[t]
    \centering
    \includegraphics[width=0.75\linewidth]{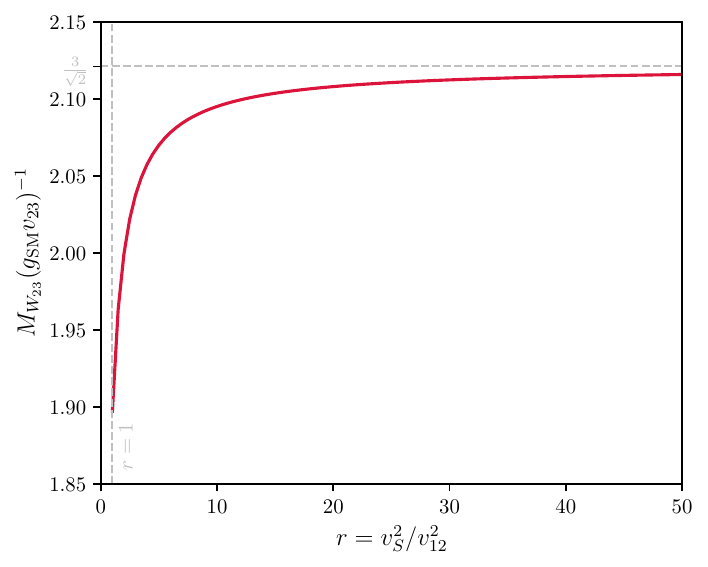}
    \caption{The mass of the $W_{23}$ as a function of $r$. The mass $M_{W_{23}}/g_\text{SM} v_{23} = 3/\sqrt{2} $  as predicted by the $\SU(2)_\text{L}^3$ model \cite{Davighi:2023xqn} is marked with the grey dashed (horizontal) line.}
    \label{fig:w23MassPlot}
\end{figure}

However, where the low scale phenomenology differs from the model with only $\SU(2)^3_\text{L}$ \cite{Davighi:2023xqn}, is the mass of the $W_{23}$ at the low scale. Due to mixing with the fields originating in the broken $\Sp(6)_\text{L}$ symmetry, $W_{23}$ has a mass which is weakly sensitive to the ratio of higher scales through the parameter $r=v_{S}^2/v_{12}^2$. Explicitly, writing the mass as $M_{W_{23}} = f(r) g_\text{SM} v_{23}$, the prefactor is given by $f(r) = \sqrt{18r/4r+1}$ which asymptotes to the value of $3/\sqrt{2}$ as predicted in the $\SU(2)^3_\text{L}$ model. This dependence is shown in Figure \ref{fig:w23MassPlot}, demonstrating this is a $\sim\mathcal{O}(10\%)$ effect at most.

\section{SMEFT Matching} \label{sec:SMEFT}
Since we have a large number of BSM fields at three distinct scales, a convenient way to consider the phenomenology of this model is by mapping onto the $d=6$ SMEFT at tree-level, using the dictionary \cite{deBlas:2017xtg}. We note that due to our chosen convention for the quark doublet $q_i = (u_i, ~ [V]_{ij} d_j)^\text{T}$ we work in the Warsaw Basis with the $u$ quarks in the mass basis \cite{Grzadkowski:2010es, Aebischer:2025qhh}. We also note that we treat the Wilson coefficients as dimensionful, with $[C] = \text{TeV}^{-2}$.

Neglecting flavour indices, the number of operators generated by this model isn't too large, because the NP of this model only couples to the the left-handed fermions, with one field also generating Higgs SMEFT operators. Therefore, at tree-level we only generate operators involving left-handed fermions or the Higgs.

Below, we show contributions to these operators in terms of the coupling matrices defined in the previous sections and the masses of the heavy gauge fields. A proper RGE treatment of the coefficients of these operators should involve computing the contribution from the field at the highest scale, then running down to the next highest-scale field and summing over contributions, repeating for all the fields of the model. The $\sum$ over fields appearing below should be understood as shorthand for this sequential matching and running procedure, though we can treat this as a naive sum if we wish for approximate coefficients which neglect running. 

\subsection{Four-Fermion Operators} 

The first operator to consider is the four-lepton operator 
\[
\mathcal{O}_{ll} = C_{ll}(\bar{l}_L \gamma_\mu l_L)(\bar{l}_L \gamma^\mu l_L),
\]
Both the BSM triplets ($W$'s and $\widetilde{W}$'s) and singlets ($Z$'s) generate this operator, with the operator coefficient 
\[
\relax[C_{ll}]_{ijkl} = \sum_{W} \left[ -\frac{\left[g_W^l \right]_{kj} \left[ g_W^l \right]_{il}}{4M_W^2} + \frac{\left[g_W^l \right]_{kl} \left[g_W^l \right]_{ij}}{8M_W^2}\right] + \sum_Z \left[- \frac{\left[g^l_Z \right]_{kl} \left[g^l_Z \right]_{ij}}{2M_Z^2} \right].
\]

Next we have the two four-quark operators 
\[
\mathcal{O}^{(1)}_{qq} = C^{(1)}_{qq} (\bar{q}_L \gamma_\mu q_L)(\bar{q}_L \gamma^\mu q_L), \qquad \mathcal{O}^{(3)}_{qq} = C^{(3)}_{qq} (\bar{q}_L \gamma_\mu \sigma^{I} q_L)(\bar{q}_L \gamma^\mu \sigma^I q_L).
\]
The massive gauge fields generate different operators depending on their SM representation. The triplets generate the $\mathcal{O}^{(3)}_{qq}$ involving $\sigma^I$ in the currents, since they have a 2-index $\SU(2)_\text{L}$ structure, whereas the singlets lacking this SU(2) structure generate the operator $\mathcal{O}^{(1)}_{qq}$. Hence the coefficients are given by
\[
\left[ C^{(1)}_{qq} \right]_{ijkl} = \sum_Z \left[ - \frac{\left[g^q_Z \right]_{kl} \left[g^q_Z \right]_{ij}}{2M_Z^2}  \right],
\]
\[
\left[ C^{(3)}_{qq} \right]_{ijkl} = \sum_W \left[ - \frac{\left[g_W^q \right]_{kl} \left[g_W^q \right]_{ij}}{8M_W^2} \right].
\]

The final four-fermion operator generated is the mixed lepton-quark operator. Again there are two of these, arising from the possible SU(2) structure of the currents, given by
\[
\mathcal{O}^{(1)}_{lq} = C^{(1)}_{lq} (\bar{l}_L \gamma_\mu l_L)(\bar{q}_L \gamma^\mu q_L), \qquad \mathcal{O}^{(3)}_{lq} = C^{(3)}_{lq} (\bar{l}_L \gamma_\mu \sigma^{I} l_L)(\bar{q}_L \gamma^\mu \sigma^I q_L).
\]
In a similar manner to the four-quark operators, the BSM triplets and singlets generate these operators separately, with the Wilson coefficients given by 
\[
\left [C^{(1)}_{lq} \right]_{ijkl} = \sum_Z \left[ - \frac{\left[g^q_Z \right]_{kl} \left[g^l_Z \right]_{ij}}{M_Z^2}  \right],
\]
\[
\left[ C^{(3)}_{lq} \right]_{ijkl} = \sum_W \left[ - \frac{\left[g_W^q \right]_{kl} \left[g_W^l \right]_{ij}}{4M_W^2} \right].
\]

\subsection{Higgs-Fermion Operators}

For left-handed fermions there are four Higgs-fermion operators of the $\mathcal{O}^{(1)}_{Hq}$, $\mathcal{O}^{(3)}_{Hq}$, $\mathcal{O}^{(1)}_{Hl}$ and $\mathcal{O}^{(3)}_{Hl}$. However, since the only field which couples to the SM Higgs is the $W_{23}$ triplet, the only operators generated at tree-level are 
\[
\mathcal{O}^{(3)}_{Hq} = C^{(3)}_{Hq} (H^\dagger \im D^I_{\text{SM},\mu} H)(\bar{q}_L \gamma^\mu \sigma^I q_L) , \qquad \mathcal{O}^{(3)}_{Hl} = C^{(3)}_{Hl}(H^\dagger \im D^I_{\text{SM},\mu} H)(\bar{l}_L \gamma^\mu \sigma^I l_L)
\]
since $\mathcal{O}^{(1)}_{Hq}$ and $\mathcal{O}^{(1)}_{Hl}$ would be generated by $Z$-type singlet fields. The Wilson coefficients then have a single contribution, and are given by
\[
\left[C^{(3)}_{Hq}  \right]_{ij} = - \frac{g^H_{W_{23}}[ g^q_{W_{23}} ]_{ij}}{4 M_{W_{23}}^2},
\]
\[
\left[C^{(3)}_{Hl}  \right]_{ij} = - \frac{g^H_{W_{23}}[ g^l_{W_{23}} ]_{ij}}{4 M_{W_{23}}^2}.
\]

\section{Phenomenology}
\label{sec:pheno}

The phenomenology of this model is predominantly dependent on the masses of the BSM gauge fields coming from the broken generators of $\Sp(6)_\text{L}/\SU(2)_\text{L,SM}$. The masses are set by the three scales of the model $v_{23}, ~ v_{12}$ and $v_S = \sqrt{r} v_{12}$, so in this section we constrain the parameter space spanned by  $(v_{23}, v_{12}, r)$.

However, there is difficulty in attempting to constrain the parameters $v_{12}$ and $r$ independently. This is a result of masses dependent on $r$ also being dependent on $v_{12}$ (see Table \ref{tab:GuageBosons}) \footnote{It should be noted that the converse of this is not true since $W_{12}$ only has $M \sim v_{12}$ (also see Table \ref{tab:GuageBosons}).}. Therefore we must  constrain the parameters $(v_{12}, r)$ simultaneously. 

Additionally observables that depend on the high scales $v_{12}$ and $v_S$ will often be dominated by their dependence on the lower scale $v_{23}$. However, since the lower scale only enters through the $W_{23}$ triplet, which has highly suppressed flavour changing couplings between the first and second generation fermions, we can isolate constraints on $v_{12}$ and $v_S$ by considering observables which involve such flavour changes. Therefore, the main observables we consider in this section are those which involve interactions between the first and second generations of fermions. Observables which involve other flavour changes (such as second to third generation) will have BSM effects dominated by the $W_{23}$ triplet and will provide constraints on $v_{23}$ analogous to those on $M_{W_{23}}$ found in \cite{Davighi:2023xqn}. Although the mass of the $W_{23}$ now has additional $r$ dependence, the mass only differs from that in \cite{Davighi:2023xqn} by $\lesssim 10\%$, so we consider constraints on $v_{23}$ briefly.

We consider observables in the presence of the new fields by first matching this model onto the SMEFT, using the Wilson coefficients defined in the previous section. Since the fields could have masses separated by several orders of magnitude, a careful analysis should take the running between mass scales into account when summing contributions to operators from different scales. The python package \texttt{Wilson} \cite{Aebischer:2018bkb} was used to handle the running during this process, which also allows translation between SMEFT and WET EFTs where necessary. In addition the python package \texttt{flavio} \cite{Straub:2018kue} was used in the computation of observables. This is the method used throughout the subsequent sections to generate the plots, though we also give approximate expressions for Wilson coefficients and observables in the absence of running.

\subsection{Meson Mixing} \label{sec:MesonMixing}

Meson observables, in particular those involving $K$ and $D$ mesons, provide powerful constraints on new physics. In the SM, any additional contributions from new particles are highly visible and typically push the NP scale to $\mathcal{O}(100\,\text{TeV})$ \cite{Isidori:2010kg}.  

The additional gauge fields of $\Sp(6)_{\text{L}}$ origin possess intrinsically off-diagonal flavour couplings so can induce meson mixing independent of the alignment between the gauge and mass bases, providing a complementary source of constraints.  

The $\Delta F = 2$ effective operators which contribute to $K^0 - \bar{K}^0$ and $D^0 - \bar{D}^0$ mixings are  
\[ \label{eq:mixingEFT}
\mathcal{L}_\text{eft} \supset - C^1_K (\bar{d}_L \gamma_\mu s_L)^2 - C_D^1(\bar{u}_L \gamma_\mu c_L)^2.
\]
These operators can be matched onto the SMEFT in the basis with $q_i = (u_i, V_{ij} d_j)^\text{T}$ by
\[
C_{D}^1 = - \left( \left[C_{qq}^{(1)}\right]_{1212} + \left[ C_{qq}^{(3)} \right]_{1212} \right) ,
\]
\[
C_{K}^1 = - (V^\dagger)_{1i} (V^\dagger)_{1k} \left( \left[C_{qq}^{(1)}\right]_{ijkl} + \left[ C_{qq}^{(3)} \right]_{ijkl} \right) V_{j2} V_{l2},
\]
where $V$ is the CKM matrix and there is an implied sum over flavour indices $i,j,k,l$. The reason for the apparent asymmetry between the two Wilson coefficients is due to our choice of (\ref{eq:q_couplings}) which contains $V_u$ in its definition, resulting in a simpler appearance for the $u$-type operator.

Using the SMEFT operators derived in §\ref{sec:SMEFT}, we can rewrite these formulae in terms of the coupling matrices of the model (\ref{eq:q_couplings}) and the masses of the heavy gauge fields by
\[ 
C_{D}^1 &= \frac{1}{8} \sum_{W} \left( \frac{\left[ g^q_{W} \right]_{12}}{M_{W}} \right)^2 + \frac{1}{2} \sum_{Z}  \left( \frac{\left[ g^q_{Z} \right]_{12}}{M_{Z}} \right)^2, \\
C_{K}^{1} &= \frac{1}{8} \sum_{W} \left( \frac{(V^\dagger)_{1i} \left[ g^q_{W} \right]_{ij} V_{j2} }{M_W} \right)^2 + \frac{1}{2} \sum_{Z} \left( \frac{(V^\dagger)_{1i} \left[ g^q_{Z} \right]_{ij} V_{j2} }{M_Z} \right)^2.
\]

\subsubsection*{$\bm{C^K_1}$}

Now we can consider the form of these operators in the various choices of alignments. Firstly, in the up-aligned scenario with $V_u=\mathbb{I}$ we have
\[ \label{eq:ck1_up_CKM}
\left(C^K_1 \right)_\text{up} = \frac{1}{16} g_{\text{SM}}^2 \Bigg[ & \frac{6 \left(V_{\text{us}} V_{\text{cd}}^* +V_{\text{cs}} V_{\text{ud}}^* \right)^2}{M^2_{\widetilde{W}_{12}}}+\frac{3 \left(V_{\text{cs}} V_{\text{cd}}^* -V_{\text{us}} V_{\text{ud}}^*\right)^2}{M^2_{W_{12}}} \\
& -\frac{6 \left(V_{\text{cs}} V_{\text{ud}}^*-V_{\text{us}} V_{\text{cd}}^*\right)^2}{M^2_{Z_{12}}}+\frac{6 \left(V_{\text{us}} V_{\text{td}}^*+V_{\text{ts}} V_{\text{ud}}^*\right)^2}{M^2_{\widetilde{W}_{13}}} \\
& -\frac{6 \left(V_{\text{ts}} V_{\text{ud}}^*-V_{\text{us}} V_{\text{td}}^*\right)^2}{M^2_{Z_{13}}}+\frac{\left(V_{\text{cs}} V_{\text{cd}}^* -2 V_{\text{ts}} V_{\text{td}}^* +V_{\text{us}} V_{\text{ud}}^ *\right)^2}{M^2_{W_{23}}} \\
& +\frac{6 \left(V_{\text{ts}} V_{\text{cd}}^*+V_{\text{cs}} V_{\text{td}}^*\right)^2}{M^2_{\widetilde{W}_{23}}}-\frac{6 \left(V_{\text{cs}} V_{\text{td}}^* -V_{\text{ts}} V_{\text{cd}}^*\right)^2}{M^2_{Z_{23}}} \Bigg]
\]
To simplify expression, we write the CKM elements in the Wolfenstein parametrisation \cite{Wolfenstein:1983yz}, which gives
\[ \label{eq:ck1_up_MW}
\left(C^K_1 \right)_\text{up} &= \frac{3}{8} g_{\text{SM}}^2 \left(\frac{1}{M^2_{\widetilde{W}_{12}}}-\frac{1}{M^2_{Z_{12}}}\right)+\frac{1}{128} \lambda ^2 g_{\text{SM}}^2 \left(\frac{96}{M^2_{W_{12}}}-\frac{192}{M^2_{\widetilde{W}_{12}}}\right) \\ 
& +\frac{1}{128} \lambda ^4 g_{\text{SM}}^2 \left(\frac{216}{M^2_{\widetilde{W}_{12}}}-\frac{96}{M^2_{W_{12}}}-\frac{24}{M^2_{Z_{12}}} \right) + \mathcal{O}(\lambda^5).
\]
noting that the higher order terms involve the other CKM parameters $A, ~ \rho$ and $\eta$. Notice that only 3 of the 8 massive gauge fields contribute at leading orders and furthermore the contribution from the $W_{23}$ (the lightest BSM field) does not appear since it is highly CKM-suppressed\footnote{Expanding the coefficient of the $W_{23}$ term in (\ref{eq:ck1_up_MW}) gives leading term $\left(V_{\text{cs}} V_{\text{cd}}^* -2 V_{\text{ts}} V_{\text{td}}^* +V_{\text{us}} V_{\text{ud}}^ *\right)^2 \approx 4 A^4 \lambda ^{10}$.} and therefore we can take this observable to be near insensitive to the scale $v_{23}$. Substituting the tree-level masses from (\ref{tab:GuageBosons}), the Wilson coefficient as a function of the model parameters $v_{12}$ and $r$ is given by
\[ \label{eq:ck1_up_v12r}
\left(C^K_1 \right)_\text{up} &= \frac{1}{v_{12}^2} \left[ \frac{1}{r (r+8)}+\lambda^2 \frac{ r-4}{8 r} - \lambda ^4\frac{ \left(r^2+4 r-36\right)}{8 r (r+8)} +\mathcal{O}( \lambda ^5 ) \right].
\]
Notice that in either of (\ref{eq:ck1_up_MW}) and (\ref{eq:ck1_up_v12r}), there is a leading term independent of the CKM parameters. This is a result of the fields $\widetilde{W}_{12}$ and $Z_{12}$ having off-diagonal couplings to the quarks before rotating from the gauge basis to mass basis with $V_u$ (see Table \ref{tab:GuageBosons}). Taking the limit of the CKM parameters going to zero ($\lambda \rightarrow 0$) in expressions (\ref{eq:ck1_up_MW}) and (\ref{eq:ck1_up_v12r}) corresponds to the choice of $V_u = V$, the \textit{down-alignment} limit. 

Hence in the \textit{down-aligned case} with $V_u = V$, or taking the CKM parameters to be zero in (\ref{eq:ck1_up_MW}) and (\ref{eq:ck1_up_v12r}), we get the expressions
\[ 
\left(C^K_1 \right)_\text{down} &= \frac{3}{8} g_{\text{SM}}^2 \left( \frac{1 }{ M^2_{\widetilde{W}_{12}}}-\frac{1 }{ M^2_{Z_{12}}} \right),
\]
\[
\left(C^K_1 \right)_\text{down} &= \frac{1}{ r (r+8) v_{12}^2}.
\]

\subsubsection*{$\bm{C^D_1}$}

Now we consider the $D$ mixing operator, $C_D^1$. This time we begin with the down-alignment case first, $V_u = V$, which yields 
\[ \label{eq:cd1_down}
\left( C_D^1 \right)_\text{down} = \frac{1}{16} g_{\text{SM}}^2 \Bigg[ &\frac{6 \left(V_{\text{us}} V_{\text{cd}}^*+V_{\text{ud}} V_{\text{cs}}^*\right)^2}{M^2_{\widetilde{W}_{12}}}+\frac{3 \left(V_{\text{ud}} V_{\text{cd}}^*-V_{\text{us}} V_{\text{cs}}^*\right)^2}{M^2_{W_{12}}} \\ 
&-\frac{6 \left(V_{\text{ud}} V_{\text{cs}}^*-V_{\text{us}} V_{\text{cd}}^*\right)^2}{M^2_{Z_{12}}}+\frac{6 \left(V_{\text{ud}} V_{\text{cb}}^* +V_{\text{ub}} V_{\text{cd}}^*\right)^2}{M^2_{\widetilde{W}_{13}}} \\
& -\frac{6 \left(V_{\text{ub}} V_{\text{cd}}^* -V_{\text{ud}} V_{\text{cb}}^* \right)^2}{M^2_{Z_{13}}}+\frac{\left(-2 V_{\text{ub}} V_{\text{cb}}^*+V_{\text{ud}} V_{\text{cd}}^*+V_{\text{us}} V_{\text{cs}}^*\right)^2}{M^2_{W_{23}}} \\
& +\frac{6 \left(V_{\text{us}} V_{\text{cb}}^*+V_{\text{ub}} V_{\text{cs}}^*\right)^2}{M^2_{\widetilde{W}_{23}}}-\frac{6 \left(V_{\text{ub}} V_{\text{cs}}^*-V_{\text{us}} V_{\text{cb}}^*\right)^2}{M^2_{Z_{23}}} \Bigg].
\]
Then again employing the Wolfenstein parametrisation we have 
\[
\left( C_D^1 \right)_\text{down} &= \frac{1}{128} g_{\text{SM}}^2 \left(\frac{48}{M^2_{\widetilde{W}_{12}}}-\frac{48}{M^2_{Z_{12}}}\right)+\frac{1}{128} \lambda ^2 g_{\text{SM}}^2 \left(\frac{96}{M^2_{W_{12}}}-\frac{192}{M^2_{\widetilde{W}_{12}}}\right) \\
&+\frac{1}{128} \lambda ^4 g_{\text{SM}}^2 \left(\frac{216}{M^2_{\widetilde{W}_{12}}}-\frac{96}{M^2_{W_{12}}}-\frac{24}{M^2_{Z_{12}}}\right)+ \mathcal{O} (\lambda ^5),
\]
Finally, inserting the tree-level masses we obtain the following expression in terms of the model parameters $v_{12}$ and $r$:
\[ \label{eq:CD1_down}
\left( C_D^1 \right)_\text{down} &= \frac{1}{v_{12}^2} \left[ \frac{1}{r (r+8)}+\lambda^2 \frac{ r-4}{8 r} - \lambda ^4\frac{ \left(r^2+4 r-36\right)}{8 r (r+8)} +\mathcal{O}( \lambda ^5 ) \right].
\]

Again as for $C_K^1$, in the opposing alignment, the Wilson coefficient would vanish if not for the off-diagonal couplings in the gauge basis coming from the broken Sp(6). In the up alignment with $V_u = \mathbb{I}$, the Wilson coefficient is 
\[
\left( C_D^1 \right)_\text{up} &= \frac{3}{8} g_{\text{SM}}^2 \left(\frac{1}{M^2_{\widetilde{W}_{12}}}-\frac{1}{M^2_{Z_{12}}}\right),
\]
or in terms of the model parameters through the tree-level masses, is given by
\[
\left( C_D^1 \right)_\text{up} = \frac{1}{r (r+8)v_{12}^2}.
\]

Notice that the formulae for $C_K^1$ and $C_D^1$ are identical in opposite alignment, even though the expressions (\ref{eq:ck1_up_CKM}) and (\ref{eq:cd1_down}) in terms of the CKM elements are different. This is a result of utilising an expansion in $\lambda$ for which the leading terms are the same.

\subsubsection*{Experimental Constraints}

In $K$ and $D$ mixing, observables are related to the off-diagonal mass matrix elements $M^K_{12}$ and $M_{12}^D$, where mass-splittings ($\Delta m_K$, $\Delta m_D$) are linearly proportional to the real part, any phase related observables ($\epsilon_K$, $q/p$) are linearly dependent on the imaginary part. In terms of the Wilson coefficients, the off-diagonal matrix elements are given by \cite{Buchalla:1995vs}
\[
M_{12}^K &= \frac{1}{2m_K} C_K^1 \braket{\bar{K}^0|(\bar{d}_L \gamma_\mu s_L)^2|K^0} \\
M_{12}^D &= \frac{1}{2m_D} C_D^1 \braket{\bar{D}^0|(\bar{u}_L \gamma_\mu c_L)^2|D^0},
\]

As we have seen in the previous section when we neglect running, to leading orders in $\lambda$ the resulting Wilson coefficients are real, and imaginary parts are only generated at higher order terms in the CKM Wolfenstein parameters\footnote{Expanding (\ref{eq:ck1_up_CKM}) and (\ref{eq:cd1_down}) to higher order in $\lambda$ finds the first imaginary terms at $\mathcal{O}(\eta A^2 \lambda^6 )$.}. Therefore even when taking running into account we expect the imaginary parts of both the Wilson coefficients and off-diagonal matrix elements to be highly suppressed compared to the real parts. 

Bounds on observables can be translated to the following constraints (at 95\% C.L) on the real parts of the Wilson coefficients \cite{SilvestriniLaThuile}
\[
C_K^1 \in &[-6.8, ~7.7] \times 10^{-13} ~ \text{GeV}^{-2}, \\
C_D^1 \in &[-2.5, ~3.1] \times 10^{-13} ~ \text{GeV}^{-2}.
\]

\begin{figure}[t]
    \centering
    \includegraphics[width=0.99\linewidth]{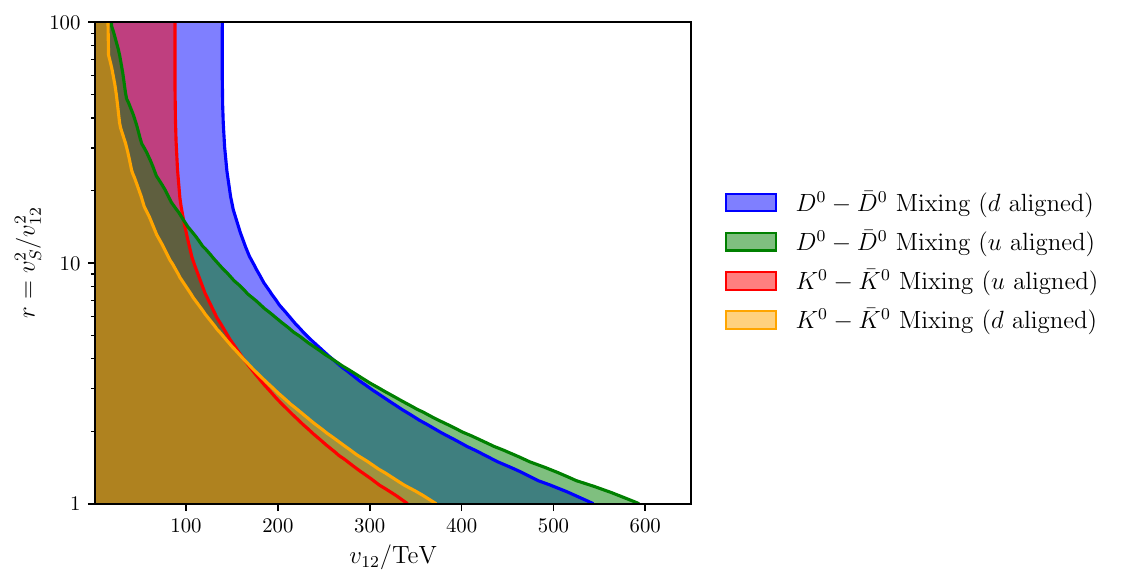}
    \caption{Exclusion on the parameter space (at $95\%$ C.L) from $K$ and $D$ meson mixing. Here $u$ and $d$ aligned refer to the choice of $V_u$ (or $V_d$) as defined in (\ref{eq:Vu_def}), with $V_u = \mathbb{I}$ referring to $u$ aligned and $V_u = V$ referring to $d$ aligned. }
    \label{fig:mesonMixing}
\end{figure}

These constraints are shown in Figure \ref{fig:mesonMixing} parameter space $(v_{12}, r)$. As expected, the constraints from $D$ mixing are much stronger than those coming from $K$ mixing which is simply a result of the constraints on the Wilson coefficient being more stringent for $D$. The curves of opposite alignment for $K$ mixing and $D$ mixing are very similar in form, a result of the leading order contributions to the Wilson coefficients being the same and the effects from running being small. 

The differences between $(C_K^1)_\text{up}$, $(C_D^1)_\text{down}$ and $(C_K^1)_\text{down}$,$(C_D^1)_\text{up}$ can be understood in terms of the formulae derived for the Wilson coefficients. Both $(C_K^1)_\text{down}$ and $(C_D^1)_\text{up}$ allow arbitrarily small $v_{12}$ for sufficiently large $r$. In terms of the EFT, this is a result of $(C_K^1)_\text{down}$ and $(C_D^1)_\text{up}$ having dependence $\sim 1/v_{S}^2$ for large $r$. Physically, this is because in these alignments the observables are only dependent on $\widetilde{W}_{12}$ and $Z_{12}$ so driving these fields to large enough masses can hide their effects. In the opposite alignments $(C_K^1)_\text{up}$, $(C_D^1)_\text{down}$ approach a bound independent of $r$ which can understood from the EFT as a result of the contribution to the Wilson coefficient $\sim \frac{\lambda^2}{v_{12}^2} $ which is independent of $r$. This is indicative of the effects from the $W_{12}$, which has mass only dependent on $v_{12}$, which is why the bound forms a cut-off on $v_{12}$.

Looking at the asymptotic behaviour as $r \rightarrow \infty$, where we decouple the $\Sp(6)/\SU(2)^3$ fields, we approach a constraint $v_{12} \gtrsim 125 ~\text{TeV}$. This is the same bound as featured in \cite{Davighi:2023xqn} though now the bound is on $v_{12}$, instead of $m_{12}$. It should be noted we approach this limit very quickly, with this constraint almost reached at $r=100$, with the scales $v_{12}$ and $v_{S}$ only separated by a factor of 10. 

One final remark is the possibility of evading the strong constraints from these observables. Since both Wilson coefficients in either alignment have a leading order term proportional to ${M^{-2}_{\widetilde{W}_{12}}}-M^{-2}_{Z_{12}}$, one could imagine having a symmetry breaking pattern which results in $M_{\widetilde{W}_{12}}=M_{Z_{12}}$, which would remove the leading order terms from $(C_K^1)_\text{up}$ and $(C_D^1)_\text{down}$ and further cause $(C_K^1)_\text{down}$ and $(C_D^1)_\text{up}$ to vanish.

\subsection{$\mu \rightarrow 3e$}
\label{sec:mu3e}

The decay $\mu^{+} \to e^{+} e^{-} e^{+}$ is a charged lepton flavour violating (CLFV) process in which a muon converts directly into three electrons without the emission of neutrinos and is highly suppressed in the SM \cite{deGouvea:2013zba, SINDRUM:1987nra}. 

In terms of the SMEFT operators generated by the fields in this model, the branching ratio can be calculated by \cite{Crivellin:2013hpa} 
\[ \label{eq:BRmu3e}
\mathrm{BR}(\mu \rightarrow 3e) = \frac{m^5_\mu}{12288 \pi^3 \Gamma_\mu} \left( 4 \left| C_{VLL}\right|^2 + |C_{VLR}|^2 + |C_{SRL}|^2 \right),
\]
where 
\[ \label{eq:WC_mu3e}
C_{VLL} &= 2 \left( (2 \sin^2\theta_W-1) \left[ C_{\phi l}^{(3)}\right]_{12} + \left[C_{ll}\right]_{1211} \right), \\
C_{VLR} &= -\frac{1}{2} C_{SRL} = 2 \sin^2\theta_W \left[ C_{\phi l}^{(3)} \right]_{12}.
\]
From the SMEFT operators derived in §\ref{sec:SMEFT} for arbitrary $V_l$, the Wilson coefficients (\ref{eq:WC_mu3e}) are given by
\[
C_{VLL} &= 2 g_{\text{SM}}^2 \Bigg(\frac{ \left(2 \sin ^2\left(\theta _{\text{W}}\right)-1\right) \left([V_l]_{11} [V_l]_{21}^*+[V_l]_{12} [V_l]_{22}^*-2 [V_l]_{13} [V_l]_{23}^*\right)}{4 M_{W^2_{23}}} \\ &-\frac{ \left([V_l]_{11} [V_l]_{11}^*+[V_l]_{12} [V_l]_{12}^*-2 [V_l]_{13} [V_l]_{13}^*\right) \left([V_l]_{11} [V_l]_{21}^*+[V_l]_{12} [V_l]_{22}^*-2 [V_l]_{13} [V_l]_{23}^*\right)}{16 M^2_{W_{23}}} \\
&-\frac{3  \left([V_l]_{11} [V_l]_{12}^*+[V_l]_{12} [V_l]_{11}^*\right) \left([V_l]_{11} [V_l]_{22}^*+[V_l]_{12} [V_l]_{21}^*\right)}{8 M^2_{\widetilde{W}_{12}}} \\
&-\frac{3  \left([V_l]_{11} [V_l]_{11}^*-[V_l]_{12} [V_l]_{12}^*\right) \left([V_l]_{11} [V_l]_{21}^*-[V_l]_{12} [V_l]_{22}^*\right)}{16 M^2_{W_{12}}} \\
&+\frac{3  \left([V_l]_{11} [V_l]_{12}^*- [V_l]_{12} [V_l]_{11}^*\right) \left( [V_l]_{11} [V_l]_{22}^*- [V_l]_{12} [V_l]_{21}^*\right)}{8 M^2_{Z_{12}}} \\
&-\frac{3  \left([V_l]_{11} [V_l]_{13}^*+[V_l]_{13} [V_l]_{11}^*\right) \left([V_l]_{11} [V_l]_{23}^*+[V_l]_{13} [V_l]_{21}^*\right)}{8 M^2_{\widetilde{W}_{13}}} \\
&+\frac{3  \left( [V_l]_{11} [V_l]_{13}^*- [V_l]_{13} [V_l]_{11}^*\right) \left( [V_l]_{11} [V_l]_{23}^*- [V_l]_{13} [V_l]_{21}^*\right)}{8 M^2_{Z_{13}}} \\
&-\frac{3  \left([V_l]_{12} [V_l]_{13}^*+[V_l]_{13} [V_l]_{12}^*\right) \left([V_l]_{12} [V_l]_{23}^*+[V_l]_{13} [V_l]_{22}^*\right)}{8 M^2_{\widetilde{W}_{23}}} \\
&+\frac{3  \left( [V_l]_{12} [V_l]_{13}^*- [V_l]_{13} [V_l]_{12}^*\right) \left( [V_l]_{12} [V_l]_{23}^*- [V_l]_{13} [V_l]_{22}^*\right)}{8 M^2_{Z_{23}}}\Bigg)
\]

\[
C_{VLR} = -\frac{1}{2} C_{SRL} = \frac{g_\text{SM}^2 \sin ^2 \theta_{\text{W}} \left([V_l]_{11} [V_l]_{21}^*+[V_l]_{12} [V_l]_{22}^*-2 [V_l]_{13} [V_l]_{23}^*\right)}{2 M^2_{W_{23}}}.
\]

For general $V_l$ it should be clear the branching ratio will be a complex expression in terms of order 4 or 8 in the elements of $V_l$ and will therefore be highly sensitive to the form of $V_l$. In an attempt to draw meaningful conclusions we consider two hierarchical forms of $V_l$ based on the two hierarchies (\ref{eq:Vl_1}) and (\ref{eq:Vl_2}) discussed previously. 

\subsubsection*{$\boldsymbol{\theta_{13}\ll\theta_{23} \ll \theta_{12}}$}

For this choice of hierarchy we take (\ref{eq:Vl}) with
\[
\theta_{12} = \alpha, \quad \theta_{23} = \alpha^2, \quad \theta_{13} = \alpha^3,
\]
where $\alpha$ is assumed to be small ($\alpha \lesssim0.1$). This choice is somewhat the most natural to consider since it mirrors the hierarchy of the CKM elements.

Under this assumption the Wilson coefficients can be expanded in powers of $\alpha$ to give 
\[
C_{VLL} = \alpha \left(\frac{3 g_{\text{SM}}^2}{2 M^2_{\widetilde{W}_{12}}}-\frac{3 g_{\text{SM}}^2}{4 M^2_{W_{12}}}\right)+\alpha ^3 \left(\frac{15 g_{\text{SM}}^2}{8 M^2_{W_{12}}}-\frac{15 g_{\text{SM}}^2}{4 M^2_{\widetilde{W}_{12}}}\right)+\mathcal{O}\left(\alpha ^5\right)
\]
\[
C_{VLR}= -\frac{1}{2} C_{SRL}  = 0 + \mathcal{O}(\alpha^5).
\]
Recalling that $C_{VLR}$ and $C_{SRL}$ are only generated at tree-level through the $W_{23}$ since it couples to the SM Higgs, the contribution from $W_{23}$ appears at $\mathcal{O}(\alpha^5)$ so we can treat the branching ratio as insensitive to $v_{23}$ for this hierarchy in the mixing angles.

Using $\Gamma_\mu = G_F^2 m_{\mu }^5/192 \pi ^3$, the branching ratio is 
\[ \label{eq:BR_1}
\text{BR}(\mu \to 3 e) = \frac{9 g_\text{SM}^4}{256 G_F^2} \Bigg[ \alpha^2 \left( \frac{ M^2_{\widetilde{W}_{12}} - 2 M^2_{W_{12}}}{M^2_{\widetilde{W}_{12}} M^2_{W_{12}}} \right)^2 - 5 \alpha^4  \left( \frac{ M^2_{\widetilde{W}_{12}} - 2 M^2_{W_{12}}}{M^2_{\widetilde{W}_{12}} M^2_{W_{12}}} \right)^2 + \mathcal{O}(\alpha^6) \Bigg],
\]
or as a function of $v_{12}$ and $r$ through substitution of the tree-level masses
\[
\text{BR}(\mu \to 3 e) = \frac{1}{1024G_F^2 v_{12}^4} \left[ \alpha^2 \frac{ (r-4)^2}{r^2}-\alpha^4 \frac{5  (r-4)^2}{r^2}+O\left(\alpha ^5\right) \right].
\]

\subsubsection*{$\boldsymbol{\theta_{13}\ll\theta_{23} = \theta_{12}}$}

In this case, we take the angles mixing $e-\mu$ and $\mu-\tau$ to be the same, with $e-\tau$ mixing suppressed:
\[
\theta_{12} = \theta_{23} = \alpha, \quad \theta_{13} = \alpha^2.
\]

With this choice of hierarchy, the $W_{23}$ is now no longer suppressed through $V_l$ and now enters the Wilson coefficients at next-to-leading order
\[
C_{VLL} &= \alpha  \left(\frac{3 g_{\text{SM}}^2}{2 M^2_{\widetilde{W}_{12}}}-\frac{3 g_{\text{SM}}^2}{4 M^2_{W_{12}}}\right) \\
&+\alpha ^3 \left(-\frac{6 g_{\text{SM}}^2}{M^2_{\widetilde{W}_{12}}}+\frac{39 g_{\text{SM}}^2}{16 M^2_{W_{12}}}+\frac{3 g_{\text{SM}}^2}{M^2_{\widetilde{W}_{13}}}+\frac{11 g_{\text{SM}}^2 \sin ^2\left(\theta _{\text{W}}\right)}{2 M^2_{W_{23}}}-\frac{55 g_{\text{SM}}^2}{16 M^2_{W_{23}}}\right)+\mathcal{O}\left(\alpha ^5\right),
\]
\[
C_{VLR} = -\frac{1}{2} C_{SRL} = \alpha ^3\frac{11  g^2 \sin ^2\left(\theta _{W}\right)}{4 M^2_{W_{23}}}+\mathcal{O}\left(\alpha ^5\right).
\]
The branching ratio is then given by
\[ \label{eq:BR_2}
\text{BR}(\mu \rightarrow 3e) = \frac{9g_\text{SM}^4}{256 G_F^2} \Bigg[&  \alpha ^2 \left( \frac{ M^2_{\widetilde{W}_{12}} - 2 M^2_{W_{12}}}{M^2_{\widetilde{W}_{12}} M^2_{W_{12}}} \right)^2\\
&+ \alpha ^4 \Bigg(\frac{16}{M^2_{\widetilde{W}_{12}} M^2_{\widetilde{W}_{13}}}-\frac{8}{M^2_{W_{12}} M^2_{\widetilde{W}_{13}}}+\frac{88 \sin ^2\left(\theta _{\text{W}}\right)}{3 M^2_{\widetilde{W}_{12}} M^2_{W_{23}}} \\
& \qquad \quad  -\frac{55}{3 M^2_{\widetilde{W}_{12}} M^2_{W_{23}}}-\frac{44 \sin ^2\left(\theta _{\text{W}}\right)}{3 M^2_{W_{12}} M^2_{W_{23}}}+\frac{55}{6 M^2_{W_{12}} M^2_{W_{23}}} \\
& \qquad \quad +\frac{29}{M^2_{\widetilde{W}_{12}} M^2_{W_{12}}}-\frac{32}{M_{\widetilde{W}_{12}}^4}-\frac{13}{2 M_{W_{12}}^4}\Bigg) + \mathcal{O}(\alpha^6) \Bigg],
\]
and using the tree-level masses the branching ratio in terms of the vevs is
\[
\text{BR}(\mu \to 3 e) &= \frac{1}{1024G_F^2 v_{12}^4} \Bigg[  \alpha^2 \frac{ (r-4)^2}{r^2}\\
&+ \alpha^4 \frac{(r-4) \left(11 (4 r+1)^2 v_{12}^2 \left(4 \cos \left(2 \theta _{\text{W}}\right)+1\right)+9 \left(-52 r^2+211 r+64\right) v_{23}^2\right)}{18 r^2 (4 r+1) v_{23}^2}+\mathcal{O}\left(\alpha ^5\right) \Bigg].
\]

Notice that the leading term is identical to that in the other choice of hierarchy. This is due to the $[V_l]_{ij}$ elements with indices $i,j \in \{1,2\}$ having the same values in either choice of hierarchy which appear in the leading terms from the contributions of $W_{12}$ and $\widetilde{W}_{12}$. We also note that in both formulae the Wilson coefficients (when ignoring running) are real, this is due to our choice to omit complex phases from $V_l$.

Furthermore, in both cases the first two terms vanish for $r\rightarrow 4$ which will heavily suppress the branching ratio. In the case of (\ref{eq:BR_1}) this can be understood as the point where $M^2_{\widetilde{W}_{12}}=2 M^2_{W_{12}}$, which from Table \ref{tab:GuageBosons} can be confirmed to be $r=4$. Although the second term of (\ref{eq:BR_2}) is more complicated, since the denominators always either contain $W_{12}$ or $\widetilde{W}_{12}$, a prefactor of $(M^2_{\widetilde{W}_{12}}-2 M^2_{W_{12}})$ arises from the combinations of the fractions and gives the same value of $r$ for cancellation.

\subsubsection*{Experimental Constraints}

The current bounds on the branching ratio come from the SINDRUM experiment \cite{ParticleDataGroup:2024cfk} and put the current bound (90\% C.L) at 
\[
\mathrm{BR}(\mu \rightarrow 3e) < 1.0 \times 10^{-12}.
\]
However, the proposed Mu3e experiment \cite{Hesketh:2022wgw} aims to increase the sensitivity by several orders of magnitude with a target of $\mathrm{BR}(\mu \rightarrow 3e) \lesssim 1 \times 10^{-16}$. In the parameter space of this model, we consider the current and (target) future bounds for hierarchies of the lepton mixing angles $\theta_{13} \ll \theta_{23} \ll \theta_{12}$ and $\theta_{13} \ll \theta_{23} = \theta_{12}$. We note that the first hierarchy is a natural choice since it mimics the CKM hierarchy seen in the quark sector, though the second hierarchy will be useful here to examine the effects of large mixings between the first and second generation charged leptons. 

\paragraph{Constraints on the mixing angle $\boldsymbol{\alpha}$.}

\begin{figure}[!ht]
    \centering
    
    \begin{subfigure}{0.99\textwidth}
        \centering
        \includegraphics[width=\linewidth]{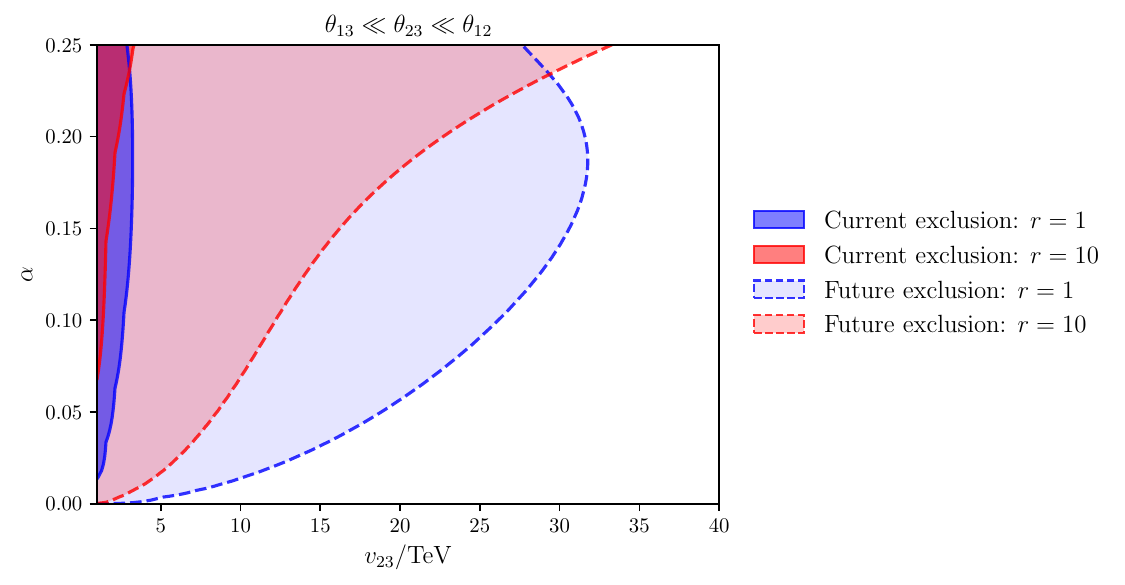}
        \caption{}
        \label{fig:alpha_ckm}
    \end{subfigure}
    
    \vspace{0.5cm}
    
    \begin{subfigure}{0.99\textwidth}
        \centering
        \includegraphics[width=\linewidth]{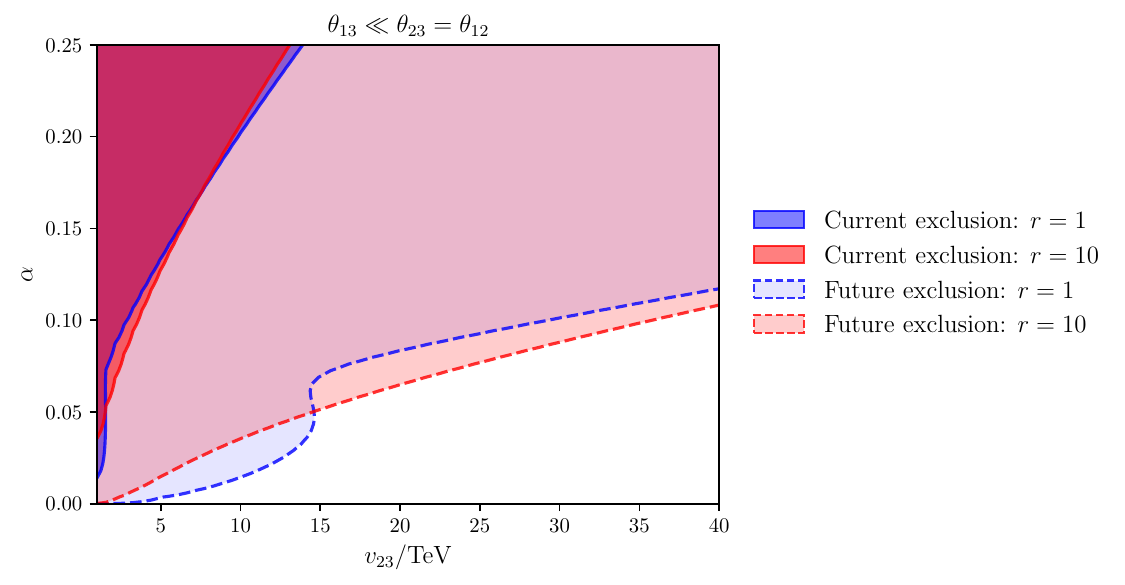}
        \caption{}
        \label{fig:alpha_other}
    \end{subfigure}
    
    \caption{Constraints on the $(v_{23}, \alpha)$ parameter space with $v_{12}=10 v_{23}$ for current bound $\text{BR} \lesssim 1 \times 10^{-12}$ \cite{ParticleDataGroup:2024cfk} and target future bound $\text{BR} \lesssim 1 \times 10^{-16}$ \cite{Hesketh:2022wgw} Figure \ref{fig:alpha_ckm} shows the current and future bounds with choice of $V_l$ and hierarchy (\ref{eq:Vl_1}), meanwhile \ref{fig:alpha_other} shows the same bounds with $V_l$ and hierarchy (\ref{eq:Vl_2}).}
    \label{fig:alphaBounds}
\end{figure}

Although we are interested in the constraints in the parameter space $(v_{12},r)$, we cannot avoid contributions coming from the low scale $v_{23}$ though the $W_{23}$ triplet. We can take this dependence as a bound on $\alpha$, dependent on $v_{23}$.

For the CKM-like hierarchy $\theta_{13}\ll \theta_{23} \ll \theta_{12}$, (\ref{eq:BR_1}) shows that the branching ratio is weakly dependent on the low scale $v_{23}$ since the contributions from $W_{23}$ enter at subleading orders. However, in the other hierarchy with somewhat large mixing between the second and third generation charged leptons, (\ref{eq:BR_2}) shows contributions from the $W_{23}$ entering at next-to-leading-order. Therefore for this hierarchy we expect strong constraints on $\alpha$ dependent on $v_{23}$.

Figure \ref{fig:alphaBounds} shows the bounds in the parameter space $(v_{23},\alpha)$ for the current bounds and target Mu3e sensitivity for the two hierarchies, with $v_{12} = 10v_{23}$. Comparing \ref{fig:alpha_ckm} and \ref{fig:alpha_other}, we can see that $\theta_{13} \ll \theta_{23} = \theta_{12}$ always imposes much stronger constraints on the mixing angle than in the case of CKM-like fermion mixing. 

In both plots of Figure \ref{fig:alphaBounds}, contours for $r>10$ do not deviate from the contour $r=10$. This is a result of the $r$-dependence here entering through the mass of the $W_{23}$. As can be seen in Figure \ref{fig:w23MassPlot}, the mass of the $W_{23}$ is asymptotic in $r$ and asymptotes for low values of $r$.

\paragraph{Constraints on $\boldsymbol{(v_{12},r)}$.}

To allow us to consider the $(v_{12}, r)$ parameter space for $\text{BR}(\mu \rightarrow 3e)$, we chose a benchmark of $v_{23} = 25\text{TeV}$. This is at a higher scale than the bounds found in \cite{Davighi:2023xqn}, though allows us to avoid artificially strong constraints coming from $\alpha$. For the CKM-like hierarchy $\theta_{13} \ll \theta_{23} \ll \theta_{12}$, we can consider $\alpha = 0.01, ~ 0.1$ which are allowed by current and future exclusions, though $r\approx1$ is excluded by the Mu3e bounds (see Figure \ref{fig:alpha_ckm}). For the other hierarchy with $\theta_{13} \ll \theta_{23} = \theta_{12}$, we use the same value of $v_{23}$, though consider angles $\alpha=0.01, ~ 0.08$. The reason for the different choices of $\alpha$ is because $\alpha = 0.1$ is excluded for all values of $r$ for $v_{23} = 25 \text{TeV}$, which can be seen on Figure \ref{fig:alpha_other}. 

\begin{figure}[!ht]
    \centering
    
    \begin{subfigure}{0.99\textwidth}
        \centering
        \includegraphics[width=\linewidth]{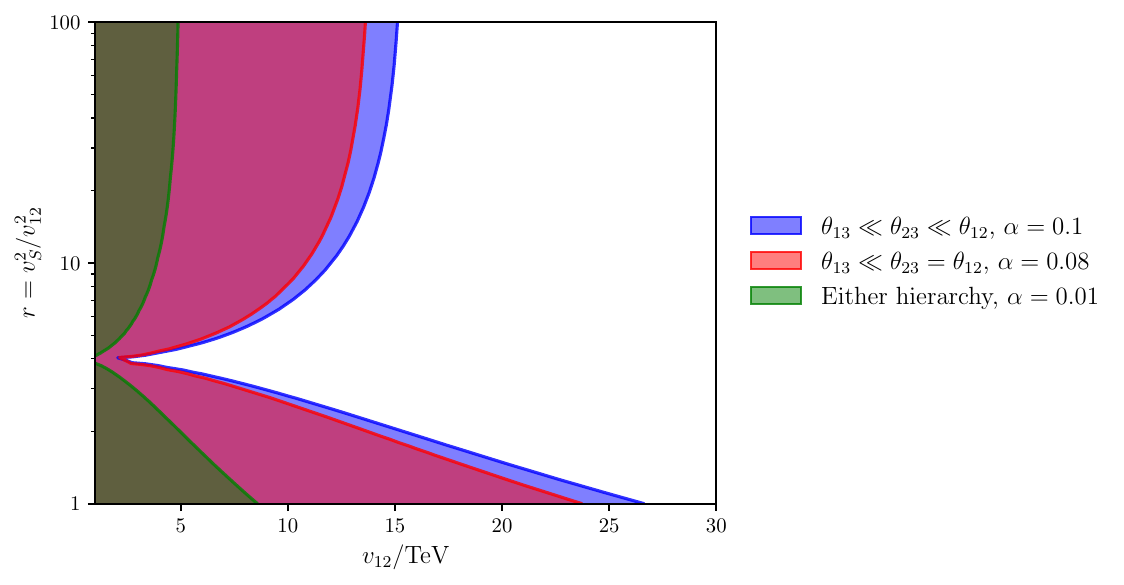}
        \caption{}
        \label{fig:Mu3e_current}
    \end{subfigure}
    
    \vspace{0.5cm}
    
    \begin{subfigure}{0.99\textwidth}
        \centering
        \includegraphics[width=\linewidth]{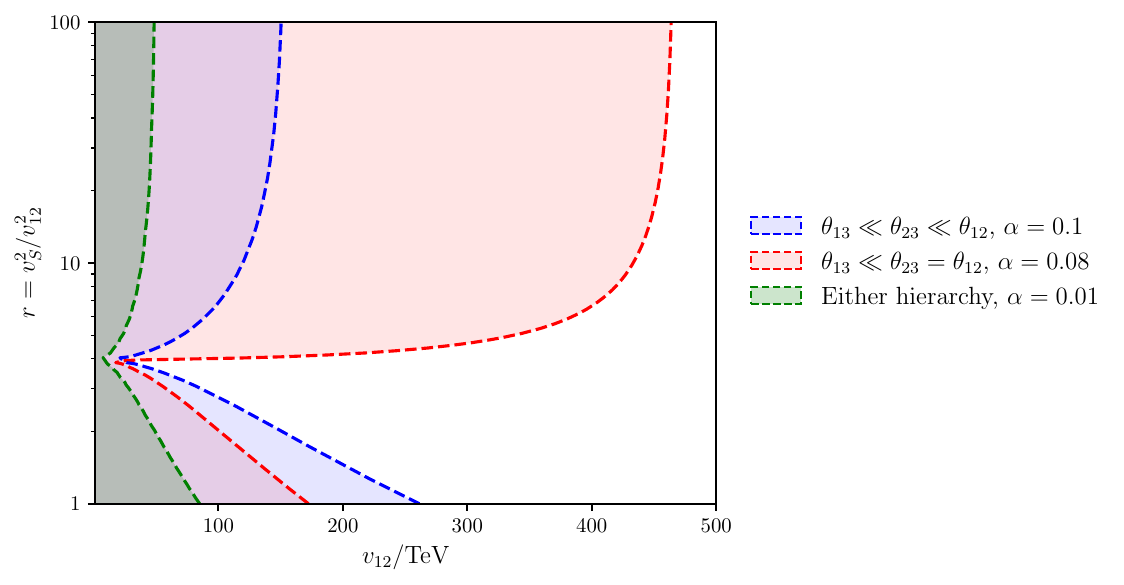}
        \caption{}
        \label{fig:Mu3e_future}
    \end{subfigure}
    
    \caption{Exclusion on the parameter space from $\text{BR}(\mu \rightarrow 3e)$. Figure \ref{fig:Mu3e_current} shows the current bound $\text{BR}<1\times10^{-12}$ \cite{ParticleDataGroup:2024cfk}, meanwhile \ref{fig:Mu3e_future} shows the target precision of the proposed Mu3e experiment $\text{BR} \lesssim 1.0 \times 10^{-16}$ \cite{Hesketh:2022wgw} All plots are drawn with $v_{23} = 25\text{TeV}$.}
    \label{fig:Mu3e}
\end{figure}

Figure \ref{fig:Mu3e_current} illustrates the current bounds as the shaded region in the parameter space $(v_{12}, r)$. As is evident from the plot, the current bounds on $\text{BR}(\mu \rightarrow 3e)$ are extremely weak when compared to meson mixing, with the bounds from meson mixing being more constraining by roughly an order of magnitude. There are still some notable features to mention. Firstly, in the $r$ direction for any choice of angle of hierarchy here, the bounds approach asymptotic values independent of $r$ for $r\approx100 $ corresponding the $v_S$ and $v_{12}$ separated by roughly one order of magnitude. The bounds approaching a limit independent of $r$ is indicative that for $r>100$ the bounds are driven by the $W_{12}$, with the effects from the $\Sp(6)/\SU(2)^3$ coset being vanishing. As $r$ becomes small, the constraints weaken, eventually allowing arbitrarily small $v_{12}$. This is a result of the fact identified earlier that the leading order contributions cancel when $2M^2_{W_{12}}=M^2_{\widetilde{W}_{12}}$, which occurs for $r=4$. Below this the bounds increase as the $\Sp(6)/\SU(2)^3$ fields become lighter until the point where the parameter $r$ is bounded. It's also worth highlighting, that the hierarchy $\theta_{13} \ll \theta_{23} = \theta_{12}$ does allow the low mass $W_{23}$ to contribute to the branching ratio at next-to-leading-order. However, since the leading contributions from the $W_{23}$ appears as a product of its mass with that of a heavier gauge field (squared), for $\alpha^4 \approx 10^{-4}$ any value of $v_{23} \gtrsim 1\text{TeV}$ results in a branching ratio within current bounds. A final comment of note is that the current bounds on the model from either hierarchy are identical, driven by the same leading contribution from $W_{12}$ and $\widetilde{W}_{12}$ in either case.

The proposed target of Mu3e provides much more stringent constraints on the parameter space, as shown in Figure \ref{fig:Mu3e_future}. The increase in sensitivity of 4 orders of magnitude results in $v_{12}$ getting driven to scales around 1 order of magnitude higher, which is expected since the branching ratio is proportional to $v_{12}^{-4}$. However, the increase in the precision also increases the sensitivity of the branching ratio to the low scale $v_{23}$.

\subsection{$\mu$ to $e$ conversion}

Muon to electron conversion describes the process in which a negatively charged muon, captured in an atomic orbital, converts directly into an electron without emitting neutrinos. This process is forbidden in the Standard Model (with only neutrino mixing it occurs at an unobservable level, $\sim 10^{-54}$), so any observation would be a clear sign of new physics~\cite{Haxton:2022piv}.  

The observable we consider is the conversion branching ratio (CR), which for a nucleus $N$ is given by
\[ \label{eq:CR}
\text{CR}(\mu \rightarrow e,~N) = 
\frac{\Gamma(\mu^{-}+N \rightarrow e^{-}+N)}{\Gamma_{\mu,\text{cap}}(N)},
\]
where $\Gamma(\mu^{-}+N \rightarrow e^{-}+N)$ is the conversion rate given by the coherent overlap of the muon and electron wavefunctions in the nuclear field, and $\Gamma_{\mu,\text{cap}}(N)$ is the experimentally known muon capture rate for nucleus $N$.  

For this model, the only contributions to the conversion rate come from the effective operators generated by the BSM gauge fields which only couple to vector currents in the left-handed fermions of the form (in the notation of \cite{Kitano:2002mt})
\[ \label{eq:mutoe_eft}
\mathcal{L}_\text{eft} \supset -\frac{G_F}{\sqrt{2}} g_{VL(q)} (\bar{e} \gamma^\mu P_L \mu)(\bar{q} \gamma_\mu q)
\]
with $q\in \{ u,d\}$. We compute the conversion rate by
\[ \label{eq:CR_width}
\Gamma(\mu^{-}+N \rightarrow e^{-}+N) = 2G_F^2 \left| \tilde{g}^{(p)}_{LV} V^{(p)} + \tilde{g}^{(n)}_{LV} V^{(n)} \right|^2
\]
where the couplings for the protons and neutrons are $\tilde{g}^{(p)}_{LV} = 2g_{LV(u)} + g_{LV(d)}$ and $\tilde{g}^{(n)}_{LV} = g_{LV(u)} + 2g_{LV(d)}$, and $V^{(p)}$ and $V^{(n)}$ are the overlap integrals which are dependent on the target nucleus. 

In terms of the SMEFT operators defined in §\ref{sec:SMEFT}, the quark Wilson coefficients are given by
\[ \label{eq:u_WC}
g_{VL,{(u)}} = -\frac{\sqrt{2}}{G_F} \left( \left[ C_{lq}^{(1)} \right]_{1211} + \left[ C_{lq}^{(3)} \right]_{1211} \right),
\]
\[ \label{eq:d_WC}
g_{VL,{(d)}} = - \frac{\sqrt{2}}{G_F} (V^\dagger)_{1k} \left( \left[C_{lq}^{(1)}\right]_{12kl} + \left[ C_{lq}^{(3)} \right]_{12kl} \right) V_{l1},
\]
where $V$ is the CKM matrix. Note that these Wilson coefficients depend on both the quark alignment through $V_u$ \textit{and} the charged lepton alignment through $V_l$. For arbitrary $V_u$ and $V_l$ we have 
\[
g_{VL{(u)}} = -\frac{g_{\text{SM}}^2}{4 \sqrt{2} G_{F}} \Bigg[ & -\frac{3 (\left| [V_u]_{11}\right|^2-\left| [V_u]_{12}\right| {}^2 ) \left([V_l]_{11} [V_l]_{21}^*-[V_l]_{12} [V_l]_{22}^*\right)}{M^2_{W_{12}}} \\
& -\frac{6 \left([V_u]_{11} [V_u]_{12}^*+[V_u]_{12} [V_u]_{11}^*\right) \left([V_l]_{11} [V_l]_{22}^*+[V_l]_{12} [V_l]_{21}^*\right)}{M^2_{\widetilde{W}_{12}}} \\
&+\frac{6 ([V_u]_{11} [V_u]_{12}^*-[V_u]_{12} [V_u]_{11}^*) ([V_l]_{11} [V_l]_{22}^*-[V_l]_{12} [V_l]_{21}^*)}{M^2_{Z_{12}}} \\
&-\frac{6 \left([V_u]_{11} [V_u]_{13}^*+[V_u]_{13} [V_u]_{11}^*\right) \left([V_l]_{11} [V_l]_{23}^*+[V_l]_{13} [V_l]_{21}^*\right)}{M^2_{\widetilde{W}_{13}}} \\
&+\frac{6 \left([V_u]_{11} [V_u]_{13}^*-[V_u]_{13} [V_u]_{11}^*\right) \left([V_l]_{11} [V_l]_{23}^*-[V_l]_{13} [V_l]_{21}^*\right)}{M^2_{Z_{13}}} \\
&-\frac{6 \left([V_u]_{12} [V_u]_{13}^*+[V_u]_{13} [V_u]_{12}^*\right) \left([V_l]_{12} [V_l]_{23}^*+[V_l]_{13} [V_l]_{22}^*\right)}{M^2_{\widetilde{W}_{23}}} \\
&+\frac{6 \left([V_u]_{12} [V_u]_{13}^*-[V_u]_{13} [V_u]_{12}^*\right) \left([V_l]_{12} [V_l]_{23}^*-[V_l]_{13} [V_l]_{22}^*\right)}{M^2_{Z_{23}}} \\
& -\frac{(\left| [V_u]_{11}\right|^2+\left| [V_u]_{12}\right| {}^2-2 \left| [V_u]_{13}\right| {}^2) \left([V_l]_{11} [V_l]_{21}^*+[V_l]_{12} [V_l]_{22}^*-2 [V_l]_{13} [V_l]_{23}^*\right)}{M^2_{W_{23}}} \Bigg], 
\]
and $g_{VL(d)}$ the same as the above expression though with $V_u \rightarrow V_d$.

When considering constraints from $\text{BR}(\mu \rightarrow 3e)$, we saw that a choice of lepton-mixing angle hierarchy $\theta_{13} \ll \theta_{23} = \theta_{12}$ lead to strong constraints even in the case of the small angles. Therefore for this observable, we only consider CKM-like charged lepton mixing. Furthermore, we also consider the extremal cases of quark alignments to simplify the expression for the Wilson coefficient. 

In the case of CKM-like charged leptons \textit{and} quarks in the up-aligned scenario
\[
\left[ g_{VL(u)} \right]_\text{up} =  -\frac{3 g_{\text{SM}}^2}{4 \sqrt{2} G_{F}} \Bigg[ -\frac{2 \alpha }{M^2_{W_{12}}}+\frac{\alpha ^3}{M^2_{W_{12}}}+\mathcal{O}\left(\alpha ^5\right)  \Bigg],
\]
\[
\left[ g_{VL(d)} \right]_\text{up} = -\frac{3 g_{\text{SM}}^2}{4 \sqrt{2} G_{F}}  \Bigg[\frac{4 \lambda }{M^2_{\widetilde{W}_{12}}}-\frac{2 \alpha }{M^2_{W_{12}}} -\frac{8 \alpha ^2 \lambda + 2\lambda^3 }{M^2_{\widetilde{W}_{12}}}+\frac{4 \alpha  \lambda ^2 + \alpha^3}{M^2_{W_{12}}}\\ +\mathcal{O}\left(\alpha^a \lambda^b; ~ a+b\geq5 \right) \Bigg],
\]
and in the opposing case with CKM-like charged leptons \textit{and} quarks in the down-aligned scenario 
\[
\left[ g_{VL(u)} \right]_\text{down} =  -\frac{3 g_{\text{SM}}^2}{4 \sqrt{2} G_{F}} \Bigg[-\frac{4 \lambda }{M^2_{\widetilde{W}_{12}}}-\frac{2 \alpha }{M^2_{W_{12}}}+\frac{8 \alpha ^2 \lambda + 2 \lambda^3 }{M^2_{\widetilde{W}_{12}}}+\frac{4 \alpha  \lambda ^2+3\alpha^3}{M^2_{W_{12}}} \\ +\mathcal{O}\left(\alpha^a \lambda^b; ~ a+b\geq5 \right)\Bigg],
\]
\[
\left[ g_{VL(d)} \right]_\text{down} =  -\frac{3 g_{\text{SM}}^2}{4 \sqrt{2} G_{F}} \Bigg[ -\frac{2 \alpha }{M^2_{W_{12}}}+\frac{\alpha ^3}{M^2_{W_{12}}}+O\left(\alpha ^5\right) \Bigg].
\]
In the cases of $\left[ g_{VL(d)} \right]_\text{up}$ and $\left[ g_{VL(u)} \right]_\text{down}$ which are dependent on both the CKM parameters and charged lepton mixing angle $\alpha$, we have expressed the Wilson coefficients as a series in combined powers of $\alpha$ and $\lambda$. In a similar manner to the other observables we consider, constraints from the conversion branching ratio are mainly due to $W_{12}$ and $\widetilde{W}_{12}$ since these contribute to processes involving flavour transitions in the first two generations without large suppression. 

One may have expected the $Z_{12}$ to also contribute at leading order here since it appeared in the other observables we have considered so far. Instead for this process, the $Z_{12}$ contribution is \textit{zero}. This arises due to the structure of its Wilson coefficient involving the CKM parameters: $([V_u]_{11} [V_u]_{12}^*-[V_u]_{12} [V_u]_{11}^*)$. In the $u$-aligned basis, this vanishes since $V_u$ is diagonal. In the $d$-aligned basis with $V_u = V$, in the Wolfenstein parametrisation $[V]_{12} = \lambda \in \mathbb{R}$. This doesn't effect the $\widetilde{W}_{12}$ since the combination of $V_u$ element does not contain a minus, as it does for the $Z_{12}$ (arising from the factors of $\im$ in the coupling matrix).

Since the nucleon Wilson coefficients are linear combinations of the above quark-level Wilson coefficients, we can never have a scenario where the nucleon Wilson coefficients, and therefore the conversion ratio (CR), are simultaneously independent of the charged lepton mixing angle and CKM parameters.

With these Wilson coefficients, the conversion rate $\Gamma$ (\ref{eq:CR_width}) to leading order in the CKM parameters and charged lepton mixing angle $\alpha$ are given by
\[
\label{eq:CR_width_up}
\Gamma_\text{up} =9g_\text{SM}^4\Bigg| \frac{ \alpha   \Delta}{2  M^2_{W_{12}}}-\frac{ \lambda    \Sigma_\text{up}}{  M^2_{\widetilde{W}_{12}}} + \mathcal{O}\left(\alpha^a \lambda^b; ~ a+b\geq3 \right)  \Bigg|^2,
\]
\[
\label{eq:CR_width_down}
\Gamma_\text{down} = 9g^4_\text{SM} \Bigg| \frac{ \alpha    \Delta}{2   M^2_{W_{12}}}-\frac{ \lambda    \Sigma_\text{down}}{  M^2_{\widetilde{W}_{12}}}+ \mathcal{O}\left(\alpha^a \lambda^b; ~ a+b\geq3 \right)\Bigg|^2,
\]
where we define the following combinations of overlap integrals
\[
\Delta = V^{(n)} - V^{(p)}, \quad \Sigma_{\text{up}} =2 V^{(n)} + V^{(p)}, \quad \Sigma_\text{down} =  V^{(n)} + 2V^{(p)}.
\]

Notice that in both cases there are leading terms which come from quark flavour transitions (dependent on $\lambda$) and charged lepton transitions (dependent on $\alpha$). Further, there is a leading term independent of the charged lepton angle $\alpha$, so we get contributions to this process even in the case $V_l = \mathbb{I}$. This should not be surprising since the process requires a $\mu - e$ transition, which is mediated by the $\widetilde{W}_{12}$ which has flavour off-diagonal couplings even when there is no mixing through $V_l$ in the charged leptons. 

Finally, as for the other observables, we can write these conversion rates in terms of $v_{12}$ and $r$ through the tree level masses, which gives 
\[
\Gamma_{\text{up}} = \frac{\alpha ^2 \Delta ^2}{16 v_{12}^4} -\frac{\alpha  \Delta  \lambda  \Sigma_\text{up}}{2 r v_{12}^4} + \frac{\lambda ^2 \Sigma_\text{up}^2}{r^2 v_{12}^4}  +\mathcal{O}\left( \alpha^a \lambda ^b; ~ a+b \geq 4\right),
\]
\[
\Gamma_\text{down} = \frac{\alpha ^2 \Delta ^2}{16 v_{12}^4} -\frac{\alpha  \Delta  \lambda  \Sigma_\text{down}}{2 r v_{12}^4} + \frac{\lambda ^2 \Sigma_\text{down}^2}{r^2 v_{12}^4}  +\mathcal{O}\left( \alpha^a \lambda ^b; ~ a+b \geq 4\right).
\]

\subsubsection*{Experimental Constraints}

\begin{figure}[htbp]
    \centering
    
    \begin{subfigure}{0.85\textwidth}
        \centering
        \includegraphics[width=\linewidth]{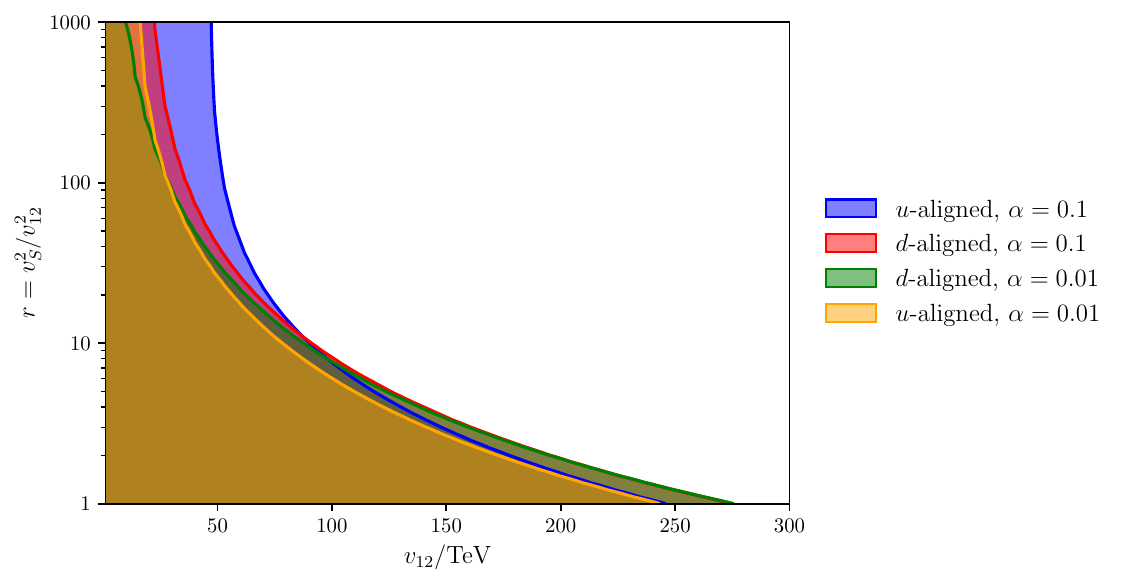}
        \caption{}
        \label{fig:CR_current}
    \end{subfigure}
    
    \vspace{0.5cm}
    
    \begin{subfigure}{0.85\textwidth}
        \centering
        \includegraphics[width=\linewidth]{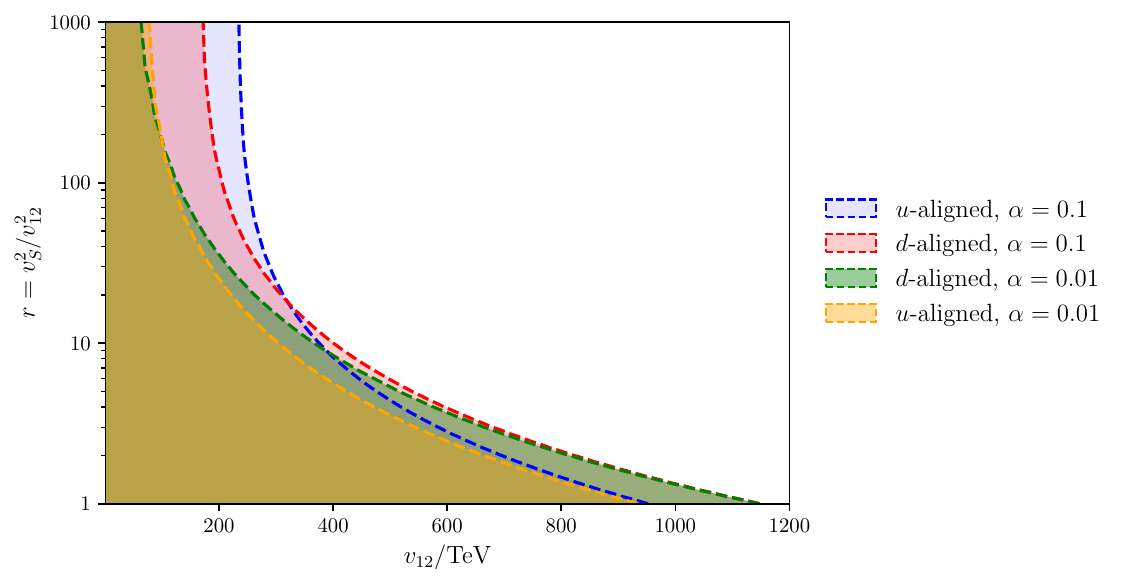}
        \caption{}
        \label{fig:CR_future}
    \end{subfigure}
    
    \caption{Exclusion on the model parameter space $(v_{12},r)$ from $\text{CR}(\mu \to e, ~N)$. Figure \ref{fig:CR_current} shows the current bounds of $\text{CR}(\mu \rightarrow e, \text{ Au}) < 7.0 \times 10 ^{-13}$ \cite{SINDRUMII:2006dvw}, and \ref{fig:CR_future} shows the projected reach of COMET-I with $\text{CR}(\mu \rightarrow e, \text{ Al}) < 1 \times 10 ^{-15}$ \cite{fujii_search_2023}. }
    \label{fig:CR_muToe}
\end{figure}

The current strongest constraints are those set by the SINDRUM-II experiment for a gold target \cite{SINDRUMII:2006dvw}:
\[
\text{CR}(\mu \rightarrow e, \text{ Au}) < 7.0 \times 10 ^{-13}.
\]
The constraints on the parameter space $(v_{12}, r)$ are shown in Figure \ref{fig:CR_current} for CKM-like mixing angles in the charged leptons with $\alpha =0.01$ and $\alpha=0.1$.

Here we see that for large $r$, the constraints from the $u$-aligned scenario always give stronger constraints than $d$-aligned, with this effect becoming more significant with increasing $\alpha$. In this regime, the constraints are dominated by effects coming from $W_{12}$ which requires $\alpha\ne0$ in order to mediate flavour changes in the charged leptons. We can see this clearly in both (\ref{eq:CR_width_up}) and (\ref{eq:CR_width_down}) as the terms $\sim \alpha^2/M^2_{W_{12}}$. 

Interestingly at the other extreme for $r\to1$, the bound is near independent of $\alpha$. This is indicative of dominating effects from $\widetilde{W}_{12}$ which allows for lepton flavour transitions even in the case there is no mixing $V_l = \mathbb{I}$, arising in both (\ref{eq:CR_width_up}) and (\ref{eq:CR_width_down}) as the terms $\sim \lambda^2$. In this region of the parameter space where $W_{12}$ and $\widetilde{W}_{12}$ have similar masses, the constraints are stronger for in the $d$-aligned case than $u$-aligned, opposite to what we see for large $r$.

The planned COMET-I experiment \cite{fujii_search_2023}, instead will utilise an aluminium target to probe a sensitivity of 
\[ \label{eq:comet-i}
\text{CR}(\mu \rightarrow e, \text{ Al}) < 1 \times 10 ^{-15}.
\]

This observable differs from that with a gold target nuclei not only by the bound being two orders of magnitude smaller, but also by the overlap integrals being smaller for aluminium \cite{Kitano:2002mt}.

Figure \ref{fig:CR_future} shows the exclusion on the parameter space from constraint (\ref{eq:comet-i}). Although the contours take a similar form to those in Figure \ref{fig:CR_current}, there is much further reach into the parameter space increasing the bounds by a factor $\sim4$.

\section{Discussion}

\begin{figure}[t]
    \centering
    
    \begin{subfigure}{0.85\textwidth}
        \centering
        \includegraphics[width=\linewidth]{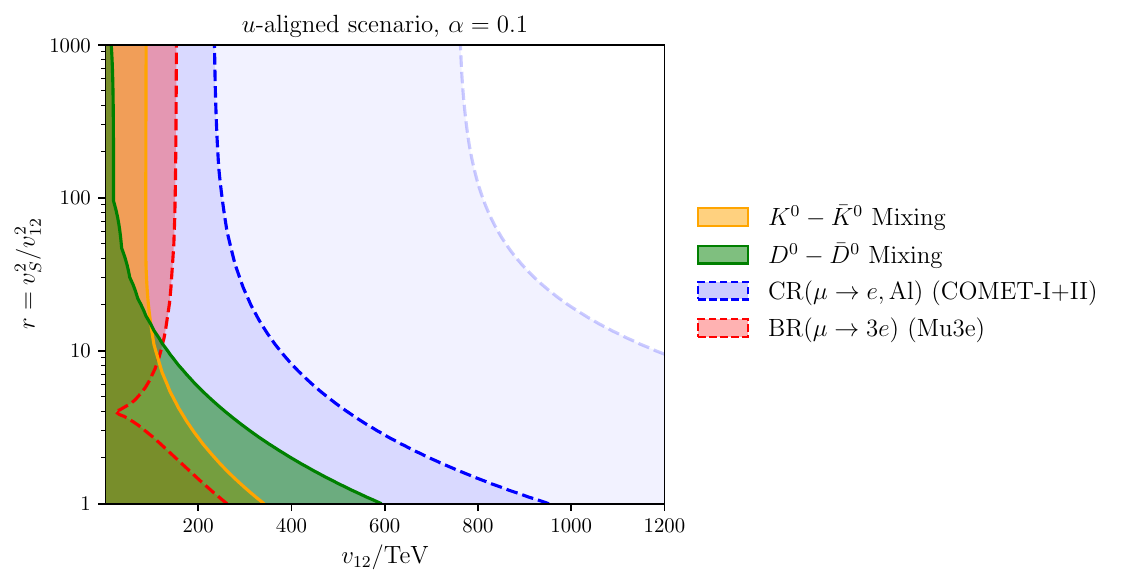}
        \caption{}
        \label{fig:comparison_up}
    \end{subfigure}
    
    \vspace{0.5cm}
    
    \begin{subfigure}{0.85\textwidth}
        \centering
        \includegraphics[width=\linewidth]{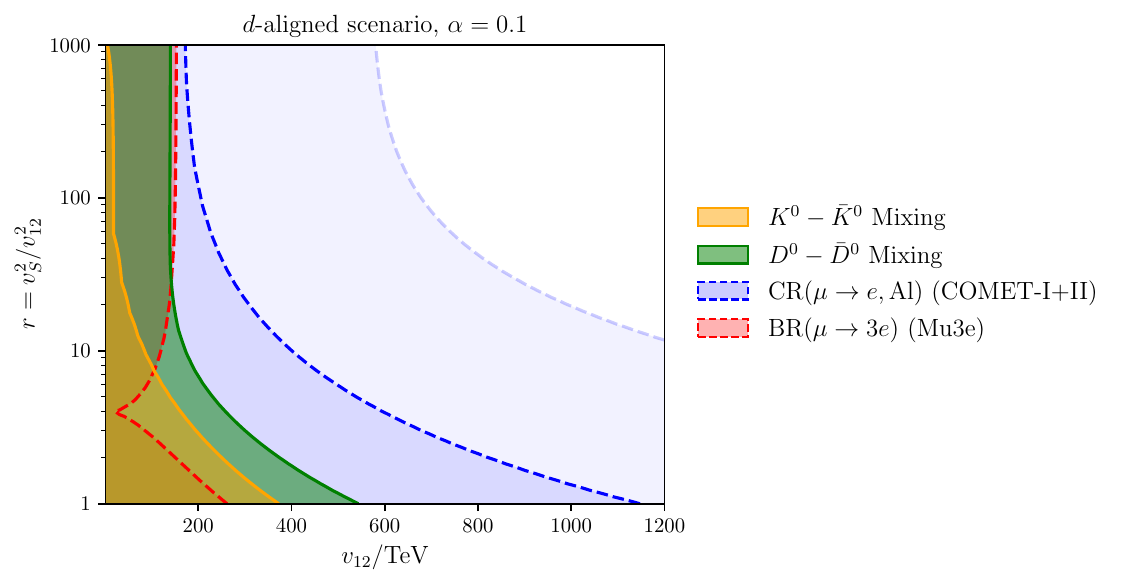}
        \caption{}
        \label{fig:comparison_down}
    \end{subfigure}
    
    \caption{Combination of strongest exclusions on the model parameter space $(v_{12},r)$ in the $u$-aligned \ref{fig:comparison_up} and $d$-aligned  \ref{fig:comparison_down} quark scenarios. In both figures solid contours represent current bounds and dashed contours represent projected reach of future experiments.}
    \label{fig:comparison}
\end{figure}

We now consider the combination of constraints on the model parameters from the flavour observables discussed in §\ref{sec:pheno}. We consider the two extremal cases of quark alignment separately, $u$-aligned constraints in Figure \ref{fig:comparison_up}, and $d$-aligned in Figure \ref{fig:comparison_down} and only consider lepton mixing of the form (\ref{eq:Vl_1}), with $\alpha =0.1$. In both plots, solid contours represent current bounds and dashed contours represent future (projected) bounds.

In either alignment the \textit{current} strongest bounds arise in meson mixing involving first and second generation quark flavour transitions. These are much stronger than the current constraints from $\mu \to 3e$ or $\text{CR}(\mu \to e , ~ \text{Au})$, in all regions of the parameter space. For $u$-alignment, the current bounds from $K^0 - \bar{K}^0$ mixing constrain the model in the large $r$ regime and $D^0 - \bar{D}^0$ mixing constrains the model for small $r$, with the bounds approximately given by
\[
\text{$u$-aligned: \quad } \begin{array}{l}
r \approx 1 \quad \implies \quad v_{12} \gtrsim 600\text{TeV} \\
r \gg 1\quad \implies \quad  v_{12} \gtrsim 100 \text{TeV}.
\end{array}
\]
In contrast to those bound, when considering the $d$-aligned scenario the FCNCs lie entirely within the $d$-type quarks, which swaps the regimes in $r$ ($r\approx1$  or large $r$) with the observables constrain. $K^0 - \bar{K}^0$ mixing constrains the model for small $r$ and $D^0 - \bar{D}^0$ constrains the model for large $r$, with the bounds roughly given by
\[
\text{$d$-aligned: \quad } \begin{array}{l}
r \approx 1 \quad \implies \quad v_{12} \gtrsim 550\text{TeV} \\
r \gg 1\quad \implies \quad  v_{12} \gtrsim 150 \text{TeV}.
\end{array}
\]
As stated in §\ref{sec:gauge-fermion}, the alignments chosen here are benchmark scenarios, since in the SM only the product $V_u V_d^\dagger = V$ is fixed by the CKM matrix. They should not be interpreted as bracketing the constraints for a general unitary $V_u$, which may contain additional CP phases. Within these benchmarks, a conservative indication of the excluded region is given by the parameter space ruled out in both alignments:
\[
\text{Current bounds ($u$-aligned + $d$-aligned): \quad } \begin{array}{l}
r \approx 1 \quad \implies \quad v_{12} \gtrsim 550\text{TeV} \\
r \gg 1\quad \implies \quad  v_{12} \gtrsim 100 \text{TeV}.
\end{array}
\]

Now turning to the prospects for future experiments, we found there is the potential to probe this model up to the 1000TeV scale. Firstly considering constraints from Mu3e projected sensitivity of the $\mu \to 3e$ branching ratio (which is independent of quark alignment) we see that there is not much improvement from the current constraints set by meson mixing. The only region of the parameter space where there is improvement is for large $r$ which sets the bound for $v_{12}$ slightly above 150TeV. It should be clear from §\ref{sec:mu3e} that the relative strength of constraints coming from the $\mu \rightarrow 3e$ process are highly dependent on the form of the charged lepton mixing matrix $V_l$. For this comparison we took the `natural' choice of $V_l$ having a similar hierarchical structure to the CKM matrix with a small parameter $\alpha$. However, we saw in §\ref{sec:mu3e} that other choices of hierarchy such as (\ref{eq:Vl_2}) can lead to much stronger constraints. Therefore, a more thorough investigation should fully explore $V_l$ by incorporating the mixing angles $\theta_{12}, ~ \theta_{23}$ and $\theta_{13}$ into the parameter space. 

Future measurements of $\mu \to e $ conversion in aluminium appear to be the most promising for probing this model, due to their potential to reach deep into the UV. From Figure \ref{fig:comparison} we can see that in either alignment the proposed sensitivity of COMET-I improves on searches into the parameter space regardless of the value of $r$. Again by looking at the regions excluded in both alignment scenarios we can make the conservative estimate that the bound set by COMET-I is
\[
\text{Future bounds (COMET-I): }\quad  \begin{array}{l}
r \approx 1 \quad \implies \quad v_{12} \gtrsim 950\text{TeV} \\
r \gg 1\quad \implies \quad  v_{12} \gtrsim 350 \text{TeV}.
\end{array}
\]
A caveat with this bound is that $\mu \to e$ conversion requires consideration of charged lepton flavour transitions. In this work we only considered $\mu \to e$ for a specific hierarchical form of the charged lepton mixing matrix $V_l$. It is very likely that for different choices of $V_l$ the constraints could either be weakened or strengthened. Therefore, similarly to $\mu \to 3e$, a more thorough consideration of $\mu \to e$ process in the context of this model would require the consideration of general lepton mixing angles present in $V_l$.

In addition to the constraints imposed by COMET-I, we also include an extra contour (lightest blue) in these plots for COMET-II \cite{Krikler:2016qij} which further aims to increase on the sensitivity of $\mu \to e$ conversion in aluminium, with a proposed sensitivity of 
\[ 
\text{CR}(\mu \rightarrow e, \text{ Al}) < 1 \times 10 ^{-17}.
\]
COMET-II demonstrates the greatest potential of upcoming experiments to probe the parameter space of this model, increasing the bounds from the already impressive COMET-I by a factor of $\sim 3$. 

From both Figures \ref{fig:comparison_up} and \ref{fig:comparison_down} we can see that small $r$ is much more restrictive on the model than for large $r$, suggesting that it is more natural for the  $\Sp(6)_\text{L}/\SU(2)^3_\text{L}$ gauge fields to be higher than the scale $v_{12}$. However, it is apparent from the plots that the separation of scales does not have to be large before constraints are independent of $r$. Almost all observables begin to show asymptotic behaviour around or before $r\approx100$, which corresponds to the $v_S/v_{12}\approx 10 $. This demonstrates that the $\text{Sp}(6)_\text{L}$ unification scale does not need to be much larger than the intermediate scale $v_{12}$ before the effects of the additional fields are hidden by the lower scale fields. Although $v_S/v_{12}\approx 10$ corresponds to $v_S \sim \mathcal{O}(1000\text{TeV})$, it demonstrates that partial unification above the SM which incorporates flavour can take place at scales not vastly beyond the SM.

Since the most sensitive experiments to new flavour physics involve the first two generations of fermions, the investigation presented here clearly favours the fields which allow unsuppressed flavour transitions, these fields being $W_{12}$, $\widetilde{W}_{12}$ and $Z_{12}$. One could consider other constraints such as those coming from $B_s$ mixing or $\tau$ LFV processes to attempt to probe the other fields of this model which mediate alternative unsuppressed flavour changes. However, as discussed in §\ref{sec:pheno}, there is difficulty in probing these fields due to the presence of $W_{23}$ at the supposed lower scale $v_{23} \sim \mathcal{O}(10\text{TeV})$ which also contributes to such processes. Therefore to consider constraints from any of the other fields (such as $\widetilde{W}_{13}, ~\widetilde{W}_{23}, ~Z_{13}$ or $Z_{23}$) one would have to do a simultaneous fit of $v_{23}$, $v_{12}$ and $r$, which was beyond the scope of this work though could be attempted in the subsequent work.

\section{Conclusions}

In the $\SU(2)^3_\text{L}$ model \cite{Davighi:2023xqn}, there were two distinct scales $v_{23}$ and $v_{12}$, which set the masses of two BSM triplet $W_{23}$ and $W_{12}$. In this work we assumed the three SU(2) factors are unified in the UV by an $\Sp(6)_\text{L}$ gauge symmetry where the three generations of doublets in the SM can be unified into a single fundamental representation (the $\vec{6}$) \cite{Davighi:2022fer}. Further, because the SM Higgs is embedded in a fundamental representation of $\Sp(6)_\text{L}$, the scalar sector necessarily contains three electroweak doublets. We have clarified in Sec.~\ref{sec:scalar_yukawa_completion} how a vacuum aligned with a single light SM-like Higgs doublet can arise from the general scalar potential, while the orthogonal doublets remain heavy non-VEV states. The same section also makes explicit that the minimal left-handed flavour embedding gives rank-1 Yukawa matrices, even after including the leading EFT operators generated by integrating out the heavy Higgs doublets. A realistic Yukawa sector therefore requires additional flavour structure, such as a larger $\SU(4)\times \Sp(6)_\text{L}\times \Sp(6)_\text{R}$ embedding or extra UV states such as vector-like fermions. We leave these scalar and Yukawa-completion questions to future work.

This enlarged symmetry group introduces an $\Sp(6)_\text{L}$ at breaking scale $v_S$ along with six more BSM fields: three triplets ($\widetilde{W}_{12}, ~ \widetilde{W}_{13}, ~ \widetilde{W}_{23}$) and three singlets ($Z_{12}, ~ Z_{13}, ~ Z_{23}$) of the SM $\SU(2)_\text{L}$. These extra fields provide sources of flavour transitions in the quarks and leptons, even in the case of quark mass-gauge basis alignment or charged lepton mass-gauge basis alignment.

We extended the results of \cite{Davighi:2023xqn} by finding the constraints on the intermediate scale $v_{12}$ in the presence of the gauge fields at the unification scale $v_S$, allowing us to map exclusions in the parameter space $(v_{12},r)$, where $r=v_S^2/v_{12}^2$. Due to the large number of new states and distinct scales involved, we adopted an effective field theory approach and matched the model onto the dimension-six Standard Model Effective Field Theory at tree-level. The resulting operator structure is highly constrained, with only operators involving left-handed fermions being generated due to this model only extending left-handed symmetries of the SM.

The constraints are dominated by the precision measurements of flavour observables involving flavour transitions between the first and second generation fermions in processes such as meson mixing ($K^0 - \bar{K}^0, ~ D^0 - \bar{D}^0$), $\mu \rightarrow 3e$ and $\mu \to e$ conversion in nuclei, since these experiments offer sensitivity to scales far beyond the direct reach of collider searches. These observables place current bounds on the intermediate scale of $v_{12}\gtrsim 550\text{TeV}$ for small $r$ when the masses of the $\Sp(6)_\text{L}/\SU(2)^3_\text{L}$ gauge fields approach the intermediate scale $v_{12}$, and $v_{12}\gtrsim 150\text{TeV}$ for large $r$ which recovers the bounds found in \cite{Davighi:2023xqn} for the $\SU(2)^3_\text{L}$ model alone. 

Overall, this work demonstrates that an $\Sp(6)_\text{L}$ origin of deconstructed weak isospin is both theoretically well-motivated and phenomenologically viable, with the model displaying a rich flavour structure. Future work could further explore $\Sp(6)_\text{L}$ or its role in the context of other models such as the promising $\SU(4)\times \Sp(6)_\text{L} \times \Sp(6)_\text{R}$ model \cite{Davighi:2022fer}.

\section*{Acknowledgments}

We are grateful to Sophie Renner for useful discussions and comments. DM is supported by the UK Science and Technology Facilities Council (STFC) under grant ST/X000605/1. AG is supported by the UK Science and Technology Facilities Council (STFC) under grant ST/X508391/1.

\bibliography{refs}
\bibliographystyle{JHEP}

\newpage
\appendix 

\section{The Sp(6) Lie Algebra}\label{appendix:sp6}

\subsection{Parametrisation of Lie Algebra Elements}

From (\ref{eq:Sp6}), the corresponding Lie algebra is defined as 
\[ \label{eq:sp6}
\mathfrak{sp}(6) \simeq \{ X \in \mathcal{M}_{6\times6}(\mathbb{C})~ | ~ X^\dagger = X, ~ \text{tr}X = 0, ~ \Omega X = - X^\text{T} \Omega  \}.
\]
The basis in which the $\SU(2)^3$ subgroup lies along the block-diagonal corresponds to a choice of 
\begin{equation} 
\Omega =  \begin{pmatrix}  
0 & 1 & 0 & 0 & 0 & 0  \\
-1 & 0 & 0 & 0 & 0 & 0  \\
0 & 0 & 0 &1 & 0 & 0  \\
0 & 0 & -1 & 0 & 0 & 0  \\
0 & 0 & 0 & 0 & 0 & 1  \\
0 & 0 & 0 & 0 & -1 & 0  \end{pmatrix}
  =  \begin{pmatrix} \epsilon & 0 & 0 \\ 0 &  \epsilon & 0 \\ 0 & 0 &  \epsilon \end{pmatrix}.
\end{equation}
Here we begin to write the $6\times6$ matrices in terms of $3\times 3$ block matrices where the elements themselves are $2\times2$ matrices. From the conditions on the elements of the Lie algebra, an element $X \in \sp(6)$ can be parametrised by
\begin{equation}
X = \begin{pmatrix} x_1 & y_{12} & y_{13} \\ y_{12}^\dagger &  x_2 & y_{23} \\ y_{13}^\dagger & y_{23}^\dagger &  x_3 \end{pmatrix},
\end{equation}
where, from (\ref{eq:sp6}), the $2\times 2$ matrices obey
\[
x_i^\dagger = x_i,\quad  \text{tr}(x_i) = 0, \quad \epsilon x_i = - x_i^\text{T} \epsilon,
\]
\[ \label{eq:y_condition}
\epsilon y_{ij} = - y_{ij}^* \epsilon.
\]
Examining the conditions on $x_i$, we see that this is precisely an element of an $\su(2)$ (where the symplectic condition is usually implicit), so we have  
\[
x_i \in \mathfrak{su}(2)_i,
\]
which is as expected since the choice of $\Omega$ aligns with $\su(2)_1 \oplus \su(2)_2 \oplus \su(2)_3$ lying along the diagonal. Therefore we can express $x_i$ in terms of $\su(2)$ generators
\[
x_i = a_1 \tau_1 + a_2 \tau_2 + a_3\tau_3 \quad \text{with} \quad a_I \in \mathbb{R}.
\]
The remaining elements $y_{ij}$ are not as constrained as the diagonal elements, with only (\ref{eq:y_condition}) constraining $y_{ij}$ to the form
\[
y_{ij} = \frac{1}{2}\begin{pmatrix}
    b_3 + \im b_4 & b_1 - \im b_2 \\ b_1 + \im b_2 & -b_3 + \im b_4
\end{pmatrix} \quad \text{with} \quad b_I \in \mathbb{R}.
\]
Notice that we can also partially write $y_{ij}$ in terms of $\su(2)$ generators, though also need to include an additional component proportional to the identity
\[ \label{eq:y_components}
y_{ij} = b_1 \tau_1 + b_2 \tau_2 + b_3\tau_3 + \frac{\im}{2} b_4\mathbb{I} \quad \text{with} \quad b_I \in \mathbb{R}.
\]
The basis matrices $\tau_I$ and $\mathbb{I}$ when embedded into the blocks $x_i$ and $y_{ij}$ form the set of generators of the $\sp(6)$ Lie algebra, which we give in the next section. The factors of $1/2$ appearing in $y_{ij}$ reflect the standard choice of generator normalisation such that $\text{tr}(T_A T_B) = \frac{1}{2}\delta_{AB}$.

Before moving on, we will introduce some useful notation for both $x_i$ and $y_{ij}$ which represent the fields present in this model. We denote the \textit{Hermitian components} with a $w$ and \textit{anti-Hermitian} components with a $z$. Therefore we write 
\[
x_i = w_{i,I} \tau_I, \quad y_{ij} = w_{ij,I} \tau_I + \frac{\im}{2} z_{ij} \mathbb{I}.
\]
For $y_{ij}$, we have the following expressions for the Hermitian and anti-Hermitian parts 
\[
y_{ij} - y_{ij}^\dagger = \im z_{ij} \mathbb{I}, \quad  y_{ij} + y_{ij}^\dagger = 2 w_{ij,I} \tau_I.
\]

\subsection{Generators}

From the previous subsection, we saw that the block elements $x_i$ form their own $\su(2)_i$ Lie algebras and that $y_{ij}$ can be written in terms of $\su(2)$ generators with an additional component proportional to the identity. We can therefore write down the generators of $\sp(6)$. In terms of $2\times 2$ blocks, the generators are given by
\[
T_{1,2,3} = \begin{pmatrix} 
    \tau_{1,2,3} & 0 & 0 \\
    0 & 0 & 0 \\
    0 & 0 & 0 \\
\end{pmatrix} &, \quad 
T_{4,5,6} = \begin{pmatrix} 
    0 & 0 & 0 \\
    0 & \tau_{1,2,3} & 0 \\
    0 & 0 & 0 \\
\end{pmatrix}, \quad 
T_{7,8,9} = \begin{pmatrix} 
    0 & 0 & 0 \\
    0 & 0 & 0 \\
    0 & 0 & \tau_{1,2,3} \\
\end{pmatrix}, \\
T_{10,11,12} = \frac{1}{\sqrt{2}} & \begin{pmatrix} 
    0 & \tau_{1,2,3} & 0 \\
    \tau_{1,2,3} & 0 & 0 \\
    0 & 0 & 0 \\
\end{pmatrix}, \quad 
T_{13} = \frac{1}{\sqrt{2}} \begin{pmatrix} 
    0 & \frac{\im}{2} \mathbb{I}  & 0 \\
    - \frac{\im}{2} \mathbb{I} & 0 & 0 \\
    0 & 0 & 0 \\
\end{pmatrix}, \\
T_{14,15,16} = \frac{1}{\sqrt{2}} & \begin{pmatrix} 
    0 & 0 & \tau_{1,2,3} \\
    0 & 0 & 0 \\
    \tau_{1,2,3} & 0 & 0 \\
\end{pmatrix}, \quad 
T_{17} = \frac{1}{\sqrt{2}} \begin{pmatrix} 
    0 & 0  & \frac{\im}{2} \mathbb{I} \\
    0 & 0 & 0 \\
    - \frac{\im}{2} \mathbb{I} & 0 & 0 \\
\end{pmatrix}, \\
T_{18,19,20} = \frac{1}{\sqrt{2}} & \begin{pmatrix} 
    0 & 0 & 0 \\
    0 & 0 & \tau_{1,2,3} \\
    0 & \tau_{1,2,3} & 0 \\
\end{pmatrix}, \quad 
T_{21} = \frac{1}{\sqrt{2}} \begin{pmatrix} 
    0 & 0  & 0 \\
    0 & 0 & \frac{\im}{2} \mathbb{I} \\
    0 & -\frac{\im}{2} \mathbb{I} & 0 \\
\end{pmatrix},
\]
where $\tau_I = \frac{1}{2} \sigma_I$, $\mathbb{I}$ is the $2\times 2$ identity and the generators are normalised to $\text{tr}(T_AT_B) = \frac{1}{2} \delta_{AB}$.

\end{document}